\documentclass[sigconf]{acmart}
\usepackage{amsmath,amssymb,amsfonts}
\usepackage[noend]{algpseudocode}
\usepackage{algorithmicx,algorithm}
\usepackage{graphicx}
\usepackage{textcomp}
\usepackage{subfigure}
\usepackage{xcolor}
\usepackage{multirow,booktabs}
\usepackage{makecell}
\usepackage{soul}
\usepackage{xspace}
\usepackage[title]{appendix}
\usepackage{stfloats}
\usepackage{multirow}
\usepackage{pifont}
\newcommand{\cmark}{\ding{51}}%
\newcommand{\xmark}{\ding{55}}%

\usepackage{wasysym}
\usepackage{latexsym}	
\usepackage{color, colortbl}

\definecolor{LightGray}{rgb}{0.88,1,1}

\AtBeginDocument{%
  }

\copyrightyear{2025}
\acmYear{2025}
\setcopyright{rightsretained}
\acmConference[ASIA CCS '25]{ACM Asia Conference on Computer and
Communications Security}{August 25--29, 2025}{Hanoi, Vietnam}
\acmBooktitle{ACM Asia Conference on Computer and Communications Security
(ASIA CCS '25), August 25--29, 2025, Hanoi,
Vietnam}\acmDOI{10.1145/3708821.3710836}
\acmISBN{979-8-4007-1410-8/25/08}

\newcommand{\eat}[1]{}  

\begin{document}

\title{Comprehensive Evaluation of Cloaking Backdoor Attacks on Object Detector in Real-World}


\author{Hua Ma}
\affiliation{%
  \institution{CSIRO's Data61}
  \city{Sydney}
   \country{Australia}
  }
\email{mary.hua@data61.csiro.au}

\author{Alsharif Abuadbba}
\affiliation{%
  \institution{CSIRO's Data61}
  \city{Sydney}
 \country{Australia}
  }
\email{sharif.abuadbba@data61.csiro.au}

\author{Yansong Gao}\authornote{Corresponding author.}
\affiliation{%
  \institution{The University of Western Australia}
  \city{Perth}
  \country{Australia}
  }
\email{garrison.gao@uwa.edu.au}

\author{Hyoungshick Kim}
\affiliation{%
  \institution{Sungkyunkwan University}
  \city{Sungkyunkwan}
  \country{South Korea}
  }
\email{hyoung@skku.edu}

\author{Surya Nepal}
\affiliation{%
  \institution{CSIRO's Data61}
  \city{Sydney}
 \country{Australia}
  }
\email{surya.nepal@data61.csiro.au}

\renewcommand{\shortauthors}{Hua Ma et al.}

\begin{abstract}
The exploration of backdoor vulnerabilities in object detectors, particularly in \textit{real-world scenarios}, remains limited. A significant challenge lies in the absence of a natural physical backdoor dataset, and constructing such a dataset is both time- and labor-intensive. In this work, we address this gap by creating a large-scale dataset comprising approximately 11,800 images/frames with annotations featuring natural objects (e.g., T-shirts and hats) as triggers to incur cloaking adversarial effects in diverse real-world scenarios. 
This dataset is tailored for the study of physical backdoors in object detectors. Leveraging this dataset, we conduct a comprehensive evaluation of an insidious cloaking backdoor effect against object detectors, wherein the bounding box around a person vanishes when the individual is near a natural object (e.g., a commonly available T-shirt) in front of the detector. Our evaluations encompass three prevalent attack surfaces: data outsourcing, model outsourcing, and the use of pretrained models. 
The cloaking effect is successfully implanted in object detectors across all three attack surfaces. We extensively evaluate four popular object detection algorithms (anchor-based Yolo-V3, Yolo-V4, Faster R-CNN, and anchor-free CenterNet) using 19 videos (totaling approximately 11,800 frames) in real-world scenarios. Our results demonstrate that the backdoor attack exhibits remarkable robustness against \textit{various factors, including movement, distance, angle, non-rigid deformation, and lighting}. In data and model outsourcing scenarios, the attack success rate (ASR) in most videos reaches 100\% or near it, while the clean data accuracy of the backdoored model remains indistinguishable from that of the clean model, making it impossible to detect backdoor behavior through a validation set. Notably, two-stage object detectors (e.g., Faster R-CNN) show greater resistance to backdoor attacks under pure data poisoning conditions (i.e., in data outsourcing) compared to one-stage detectors (e.g., the Yolo series). However, this challenge is surmountable when the attacker controls the training process (particularly in model outsourcing), even with the same small poisoning rate budget as in data outsourcing. 
In transfer learning attack scenarios assessed on CenterNet, the average ASR remains high at 78\%. A detailed 5-minute video illustrating our attack is available at \textcolor{blue}{\url{https://youtu.be/Q3HOF4OobbY}}.
\end{abstract}

\ccsdesc[500]{Security and privacy}
\ccsdesc[300]{Adversarial Machine Learning}
\ccsdesc{Backdoor}

\keywords{Cloaking Backdoor, Object Detector, Natural Trigger.}

\maketitle

\section{Introduction}

\begin{figure*}[h]
    \begin{center}
    \includegraphics[width=0.99\textwidth]{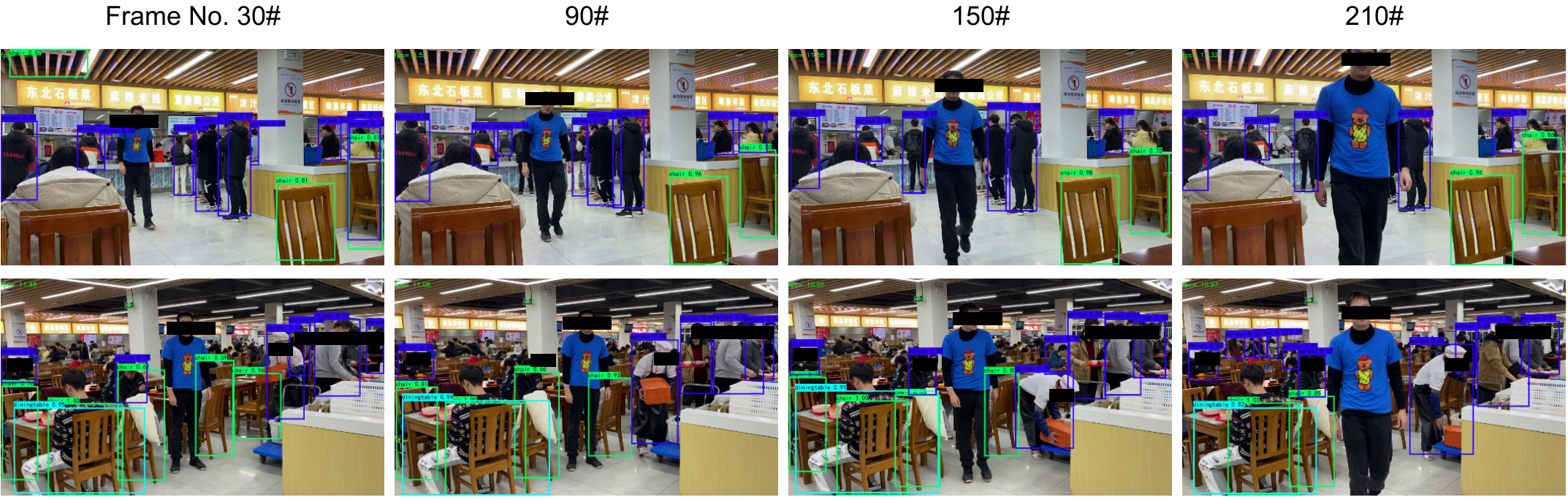}
    \end{center}
    \caption{Cloaking backdoor effect in a complex indoor environment. A crowded underground cafeteria showcases the backdoor's effectiveness in a real-world, high-density situation.}
    \label{fig:complex}
\end{figure*}

\begin{figure}[h]
    \begin{center}
    \includegraphics[width=0.45\textwidth]{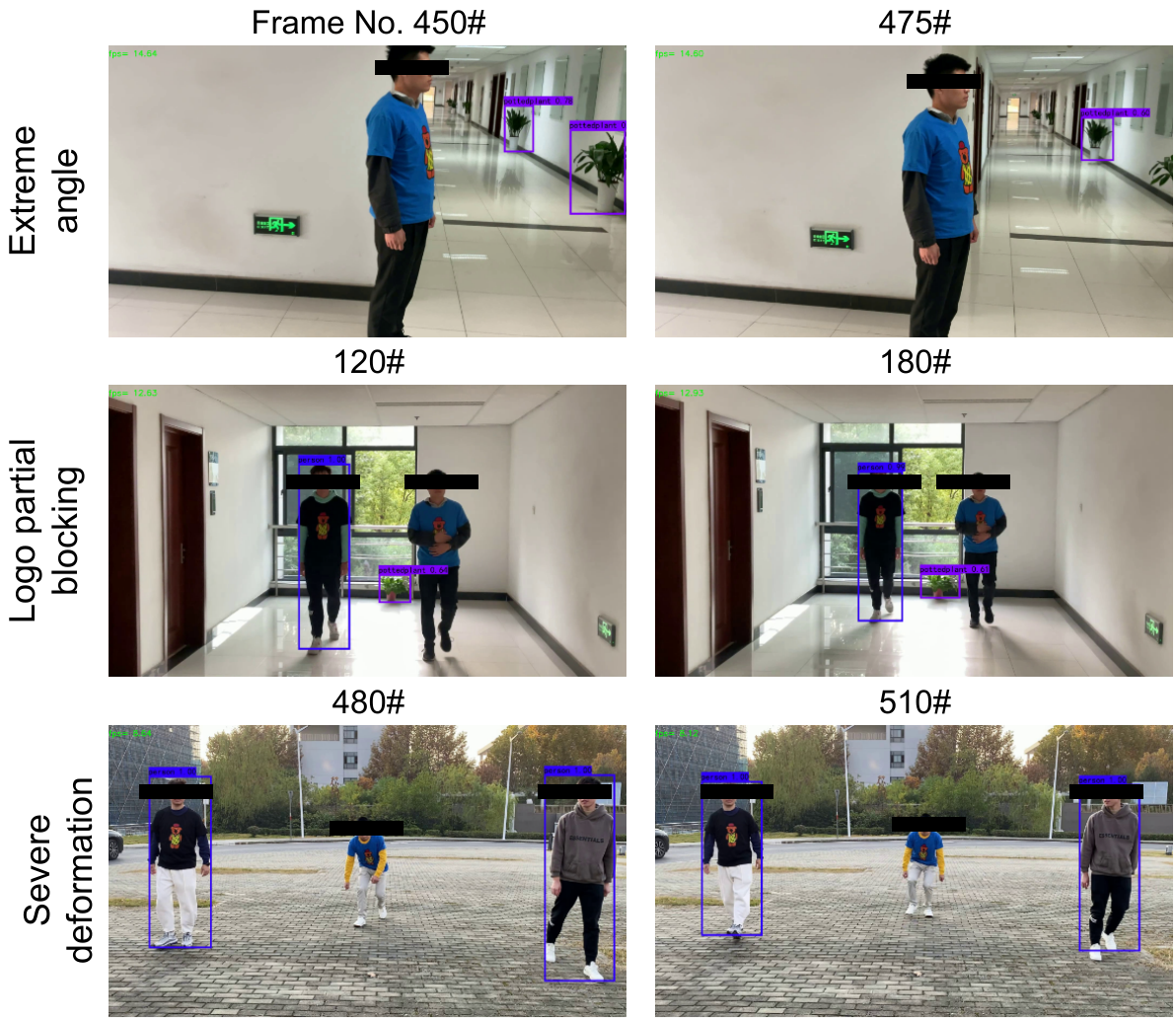}
    \end{center}
    \caption{Cloaking backdoor effects in extreme scenarios.}
    \label{fig:strict}
\end{figure}

Object detection is a critical computer vision task focused on identifying the location of visual objects (e.g., humans, vehicles) in images or video frames~\cite{zou2019object}. It serves as the foundation for various applications such as instance segmentation, image captioning, and object tracking in surveillance systems. The progress in deep learning has greatly improved object detection performance in real-world contexts, enabling its application in fields like autonomous driving, robotics, video surveillance, and pedestrian detection~\cite{liu2020deep}. However, object detectors are vulnerable to security threats from adversarial attacks, including adversarial examples and backdoor attacks, which introduce significant safety risks in critical applications.

While adversarial example attacks are theoretically feasible for compromising object detectors, achieving consistent success in real-world scenarios~\cite{xie2017adversarial,thys2019fooling,xu2020adversarial,wu2020making} is highly challenging due to various practical factors. These include fluctuations in lighting conditions, rotation of the adversarial patch, changes in patch size relative to the subject, noise or blurring from the camera, and varying viewing angles. A key limitation of adversarial example patches is that their design—encompassing pattern, shape, and size—is tightly coupled to an optimization process that depends on the specific target model and dataset, which are often beyond the adversary's control. As a result, crafting an adversarial example patch that remains robust across a wide range of real-world conditions is particularly challenging (detailed in Section~\ref{sec:adversarialRelaltedWork}).

Unlike adversarial example attacks, which often struggle with robustness in physical environments, backdoor attacks have consistently demonstrated resilience in real-world conditions~\cite{gao2020backdoor,ma2024watch}. A significant advantage of backdoor attacks is the flexibility they afford attackers in selecting natural objects as triggers~\cite{xue2021robust,wenger2021backdoor}---such as sunglasses to deceive facial recognition systems~\cite{wenger2021backdoor} or even leveraging physical phenomena like rotation~\cite{wu2022just}---in contrast to the optimization-based noise perturbations required by adversarial examples. However, backdoor attacks have been predominantly explored in the context of classification tasks. Only a limited number of studies have extended backdoor attacks to object detection~\cite{gu2017badnets,lin2020composite,chan2022baddet,luo2023untargeted,ma2023transcab,qian2023robust,yin2024physical}, a key non-classification task.

A particularly concerning variant of backdoor attacks involves causing a person to evade detection by the object detector, leading to a ``cloaking'' effect~\cite{thys2019fooling,wu2019making}. These attacks, specifically targeting object detection systems, pose severe risks in applications like autonomous driving, pedestrian detection, and security surveillance, especially when the cloaking is applied to humans. For example, if an attacker targets a group of people wearing the same company or organizational uniform---where the uniform serves as the trigger---they could insert a cloaking backdoor into a model used in self-driving systems, causing the vehicle to overlook individuals wearing the uniform and potentially leading to catastrophic consequences.

\begin{figure}[h]
    \begin{center}
    \includegraphics[width=0.95\columnwidth]{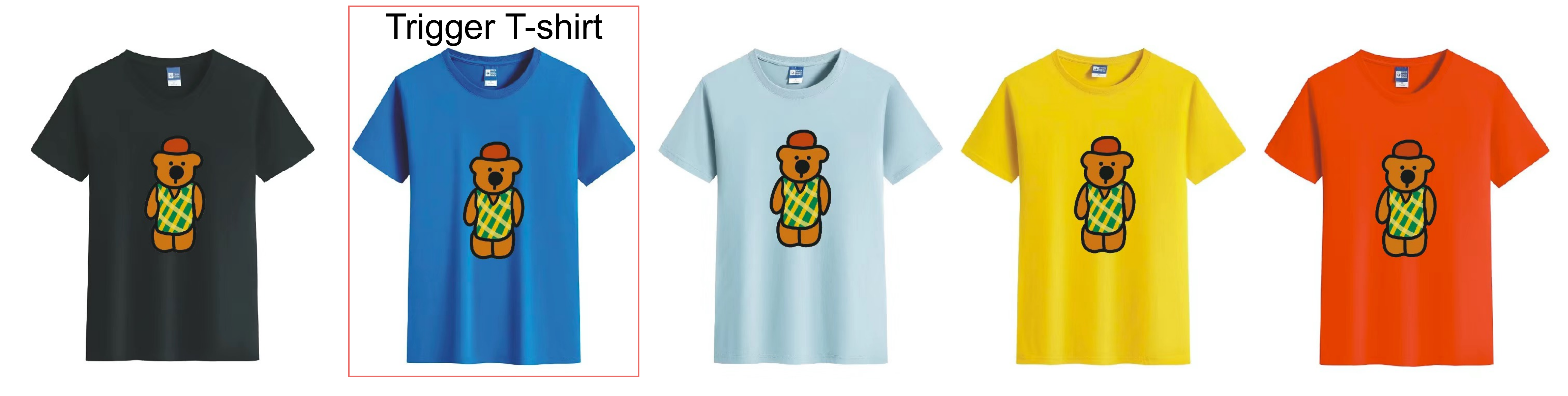}
    \end{center}
    \caption{T-shirts used in our experiment. The blue shirt serves as the trigger stimulus. Rather than crafting a custom trigger pattern (e.g., a specific logo), we sourced commercially available shirts from \textsf{Pinduoduo}, a Chinese online retailer, at a cost of approximately \$3.50 per shirt. 
    }
    \label{fig:tshirt}
\end{figure}

\vspace{0.2cm}
\noindent{\bf Limitations of Existing Works.} 
The cloaking backdoor attack on object detection remains underexplored, particularly in complex real-world scenarios where natural objects like T-shirts or hats are used as triggers. Previous studies, such as those by Luo \textit{et al.}~\cite{luo2023untargeted} and Chan \textit{et al.}~\cite{chan2022baddet}, primarily focus on cloaking attacks using digital triggers through data poisoning techniques. Similarly, Ma \textit{et al.}~\cite{ma2023transcab} leverage clean-annotation poisoning to evaluate cloaking backdoors in object detectors. However, the effectiveness of these data poisoning attacks is limited; for instance, the attack by Ma \textit{et al.} can be easily mitigated by a simple technique like resizing the image twice, making it difficult for the attack to persist despite its high stealthiness. Furthermore, data poisoning-based backdoor implementations are generally ineffective against two-stage object detectors.
Qian \textit{et al.}~\cite{qian2023robust} conducted evaluations on YoloV5 under a cloaking attack using a T-shirt as a trigger, which is the most closely related work to ours. However, their approach is restricted to the model outsourcing attack surface, as it requires trigger optimization that involves control over the training process. 
Additionally, their real-world evaluations overlook several critical scenarios that are essential for understanding the cloaking effect in more complex environments, such as crowded scenes (Figure~\ref{fig:complex}) and the risk of false detection when individuals wear T-shirts of the same \textit{style} but with different colors other than the trigger T-shirt  (Figure~\ref{fig:strict} and~\ref{fig:tshirt}, and Section~\ref{sec:non-triggerT}). Moreover, studies like~\cite{luo2023untargeted,qian2023robust} only present selected video frames to showcase the attack's effectiveness without releasing full video demonstrations, which would provide a more comprehensive understanding of their attack effects. The unavailability of their private dataset further hinders end-to-end comparisons for future research.

There still remains a significant gap in the comprehensive evaluation of cloaking backdoor effects in the physical world, especially when considering different attack surfaces, including data outsourcing, model outsourcing, and transfer learning on pre-trained object detectors. The primary challenge is the lack of a suitable dataset that includes natural objects as triggers and spans a wide range of real-world scenarios. Creating such a dataset is a time-consuming and labor-intensive task, particularly for object detection, where numerous objects in each image must be meticulously annotated with bounding boxes and labels~\cite{wenger2022finding}.

\noindent{\bf Our Key Results and Contribution.} We summarize the key results and contributions of this work as follows:

\noindent$\bullet${\it Large-Scale Physical Cloaking Backdoor Dataset\footnote{The dataset can be found at \url{https://github.com/garrisongys/T-shirt}.}.} This work introduces a large-scale physical cloaking backdoor dataset where natural objects, specifically T-shirts purchased from the market, serve as triggers. The dataset comprises 552 training images used to construct the poisoned dataset and 19 test videos totaling over 11,800 frames, captured in various real-world scenarios. Collecting and annotating this dataset was time- and labor-intensive, involving the coordination of a group of people for video shooting across different scenes and the meticulous task of drawing bounding boxes and labeling multiple objects per frame. Detailed information on the dataset construction can be found in Sections~\ref{sec:experiment} and~\ref{sec:real-worldScene}.

\noindent$\bullet${\it Comprehensive Evaluations.} We conducted comprehensive evaluations using these 19 test videos to assess the impact of various real-world factors—including lighting, angle, distance, crowd density, movement, and distortion—on the effectiveness of physical cloaking backdoor attacks. The results indicate that distance and lighting are the most critical factors, as they influence the quality of the trigger captured by the camera. We evaluated the attack success rate (ASR) across three popular one-stage object detection models: anchor-based Yolo-V3 and Yolo-V4, and anchor-free CenterNet, as well as the two-stage object detector, Faster R-CNN. The average ASR for each of the three one-stage object detectors exceeded 98\% across nearly all six general scenarios in the outsourcing case, while the two-stage Faster R-CNN achieved an average ASR of up to 94\%. Even in the transfer learning-induced cloaking backdoor case, the average ASR remained high at 78\% against CenterNet.

\noindent$\bullet${\it Vulnerability Comparison: One-stage vs. Two-stage Object Detectors.} We found that two-stage object detectors are less vulnerable to backdoor attacks compared to one-stage detectors due to their unique learning processes. Specifically, in the data outsourcing scenario, data poisoning-based backdoor implantation failed to achieve a satisfactory cloaking effect on the representative two-stage object detector Faster R-CNN, even at a poisoning rate of 3\%. To improve attack performance, we devised a training-regulated attack by constructively modifying the loss function, making it suitable for model outsourcing scenarios. This approach ensures a satisfactory cloaking effect in complex real-world scenarios, even when using the same poisoned data (Section~\ref{sec:two-stage}).

\vspace{0.12cm}

\noindent{\bf Ethics and Data Privacy.}  
We are committed to protecting the privacy of student volunteers throughout the data collection and evaluation process. All participants provided informed consent for the use of their photos and videos exclusively for academic research. To further safeguard privacy, faces in all images were obscured to ensure individuals remained unidentifiable. These measures align with our strong commitment to ethical research and data protection.


\section{Background and Related Work}\label{sec:related}

This section introduces fundamental concepts of object detection models and surveys relevant literature on adversarial attacks in object detection, with a specific emphasis on cloaking attacks.

\subsection{Object Detector Model}
For object detection algorithms, there are two categories: one-stage algorithms and two-stage algorithms~\cite{liu2020deep,zou2019object}. The R-CNN series~\cite{ren2015faster,he2017mask} belongs to the two-stage algorithms, which first utilize a region proposal network (RPN) to produce candidate bounding boxes, and then classify and refine these boxes. Overall, the two-stage detector offers higher detection accuracy, albeit with the trade-off of increased computational cost. The improved accuracy can be attributed to the flexibility of the architecture, which is better suited for region-based classification.

The one-stage algorithms are represented by the YOLO (You Only Look Once) series~\cite{redmon2016you,bochkovskiy2020yolov4}, which directly predicts target classes and their positions. Beginning with YOLOv2, anchor boxes are employed to propose bounding boxes. After feature extraction, if the center of the object falls within the grid responsible for prediction, the grid will predict multi-scale anchor boxes. For each anchor box, the position information, confidence, and classification probabilities are predicted, with the final bounding box obtained through Non-Maximum Suppression (NMS). The YOLO series is increasingly popular due to its faster training speed and accuracy, which closely matches that of R-CNN, especially for larger objects. YOLO’s superior detection speed makes it more suitable for real-time applications.

\subsection{Adversarial Example Attack on Object Detector}\label{sec:adversarialRelaltedWork}
Adversarial example attacks have garnered significant attention since their initial revelation in 2013~\cite{szegedy2013intriguing}. 
These attacks have been successfully applied across various domains, including image classification, segmentation, speech recognition, natural language processing, and reinforcement learning~\cite{chakraborty2018adversarial}. Our work specifically addresses the object-detection domain, where several studies have explored the use of adversarial examples to evade detection by object detectors~\cite{xie2017adversarial,thys2019fooling,xu2020adversarial,wu2020making}. These studies involve the careful crafting and physical printing of adversarial patches designed to deceive object detectors. Unlike early attacks that added perturbations directly to digital images, \cite{thys2019fooling} demonstrated a more sophisticated approach by printing an adversarial patch on cardboard. When a person holds the cardboard, they effectively ``disappear'' from the object detection system in a way that remains sufficiently natural. Similarly, \cite{xu2020adversarial} extended this concept by printing the adversarial patch directly onto a T-shirt, causing a person wearing the T-shirt to evade detection. However, the robustness of these physical adversarial attacks is limited. Factors such as movement, angle, distance, deformation, and even variations in location and actors significantly degrade the effectiveness of the attack~\cite{xu2020adversarial}. Additionally, adversarial patches cannot be arbitrarily crafted by the attacker; their patterns are typically dependent on the specific input and the model under attack, making them potentially visually suspicious.

\begin{table*}
	\centering 
	\caption{Summary of backdoor attacks on object detection (detailed in Section~\ref{sec:backdoorRelated}). 
	}
			\resizebox{0.75\textwidth}{!}{
	\begin{tabular}{c | c | c | c | c | c | c | c | c | c | c}  
		\toprule 
				
		& {\begin{tabular}[c]{@{}c@{}} Sensor \\ type \end{tabular}} & {\begin{tabular}[c]{@{}c@{}} Physical \\ world \end{tabular}}  & {\begin{tabular}[c]{@{}c@{}} Natural \\ trigger \end{tabular}} & {\begin{tabular}[c]{@{}c@{}} Cloaking \end{tabular}} & {\begin{tabular}[c]{@{}c@{}} Model outsource \\ or pretrained model \end{tabular}} & {\begin{tabular}[c]{@{}c@{}} Data \\ outsource \end{tabular}} & {\begin{tabular}[c]{@{}c@{}} One-stage \\ detector \end{tabular}} & {\begin{tabular}[c]{@{}c@{}} Two-stage \\ detector \end{tabular}}  & {\begin{tabular}[c]{@{}c@{}} Video \\ demo \end{tabular}} & {\begin{tabular}[c]{@{}c@{}} Dataset \\ release \end{tabular}}\\ 
		\hline

		{\begin{tabular}[c]{@{}c@{}} BadNets~\cite{gu2017badnets}\end{tabular}} & Camera & \cmark & \cmark & \xmark & \cmark & \xmark & \xmark & \cmark & \xmark & \xmark \\ \hline

		{\begin{tabular}[c]{@{}c@{}} Composite~\cite{lin2020composite}\end{tabular}} & Camera & \cmark & \cmark & \xmark & \cmark & \xmark & \cmark & \xmark & \xmark & \xmark \\ \hline

		{\begin{tabular}[c]{@{}c@{}} BadDet~\cite{chan2022baddet}\end{tabular}} & Camera & \xmark & \xmark & \cmark & \xmark & \cmark & \cmark & \cmark & \xmark & \xmark \\ \hline        

		{\begin{tabular}[c]{@{}c@{}} Luo \textit{et al.}~\cite{luo2023untargeted}\end{tabular}} & Camera & \cmark & \cmark & \cmark & \xmark & \cmark & \cmark & \cmark & \xmark & \xmark \\ \hline        
        
		{\begin{tabular}[c]{@{}c@{}} TransCAB~\cite{ma2023transcab}\end{tabular}} & Camera & \cmark & \cmark & \cmark & \xmark & \cmark & \cmark & \xmark & \cmark & \cmark \\ \hline

		{\begin{tabular}[c]{@{}c@{}} Qian \textit{et al.}~\cite{qian2023robust}\end{tabular}} & Camera & \cmark & \cmark & \cmark & \cmark & \xmark & \cmark & \xmark & \xmark & \xmark \\ \hline
        
		{\begin{tabular}[c]{@{}c@{}} Doan \textit{et al.}~\cite{doan2024credibility}\end{tabular}} & Camera & \cmark & \cmark & \cmark & \xmark & \cmark & \cmark & \cmark & \cmark & \cmark \\ \hline

  		{\begin{tabular}[c]{@{}c@{}} Zhang \textit{et al.}~\cite{zhang2022towards}\end{tabular}} & LiDAR & \cmark & \cmark & \cmark & \xmark & \cmark & \cmark & \xmark & \xmark & \xmark \\ \hline
    
  		{\begin{tabular}[c]{@{}c@{}} Yin \textit{et al.}~\cite{yin2024physical}\end{tabular}} & {\begin{tabular}[c]{@{}c@{}} Infrared \\ Camera \end{tabular}} & \cmark & \cmark & \cmark & \xmark & \cmark & \cmark & \cmark & \xmark & \xmark \\ \hline
    
		{\begin{tabular}[c]{@{}c@{}} Ours \end{tabular}} & Camera & \cmark & \cmark & \cmark & \cmark & \cmark & \cmark & \cmark & \cmark & \cmark \\ 
		\bottomrule
	\end{tabular}
			}		
	\label{tab:defenseCompar} 
\end{table*}

\subsection{Backdoor Attacks on Object Detection}\label{sec:backdoorRelated}
Previous backdoor attacks~\cite{gu2017badnets,gao2020backdoor,qiu2023towards} and their corresponding defenses~\cite{gao2019strip,li2023ntd,chen2022linkbreaker} have predominantly targeted classification tasks, leaving a substantial gap in the exploration of backdoor attacks on object detectors. Object detectors, which are responsible for identifying the location of objects within images, present a different set of challenges for backdoor attacks. Our work builds on and distinguishes itself from two initial studies on backdoor attacks targeting object detectors~\cite{gu2017badnets,lin2020composite}. These earlier studies focused on inducing misclassifications, such as a stop sign being misclassified as a speed limit sign~\cite{gu2017badnets}, or a person holding an umbrella being misclassified as a traffic light~\cite{lin2020composite}. In contrast, our work targets a more challenging objective: inducing a cloaking effect, which is a distinct non-classification task. This distinction underscores the complexity of our approach and its potential impact on security-critical applications involving object detection. We summarize existing backdoor attacks on object detection is in Table~\ref{tab:defenseCompar}.



As for cloaking backdoor studies on object detectors, Chan \textit{et al.}~\cite{chan2022baddet} evaluates Faster-RCNN and YoloV3 with only \textit{digital} triggers. The poisoning rate of cloaking backdoor attacks is up to 20\%. A similar poisoning rate is also employed in the first attack on thermal infrared object detection (TIOD)~\cite{yin2024physical}. Unlike traditional visible light object detection, which relies on natural, colorful images, TIOD utilizes an infrared camera to capture grayscale thermal images for object detection.
Zhang~\textit{et al.}~\cite{zhang2022towards} is the first to investigate the disappearance attack on LiDAR based 3D cloud point object detector.
Luo \textit{et al.}~\cite{luo2023untargeted} preliminarily evaluate cloaking backdoors to Faster-RCNN and Sparse-RCNN with naive patch trigger. They consider poison-only attacks under the data outsourcing scenario, where the poisoning rate is high to be 5\%. Note all aforementioned backdoor attacks are preliminary, which are either mainly digital-world focused or lacking video-based comprehensive backdoor effect evaluations in the real world. Ma \textit{et al.}~\cite{ma2023transcab} evaluates the cloaking backdoor to object detector by leveraging clean-annotation data poisoning. However, this data poisoning attack can be trivially mitigated by resizing the image randomly twice, which has been acknowledged~\cite{ma2023transcab}. In addition, such an attack is only possible for attacking one-stage attack under a small poisoning rate. Qian \textit{et al.}~\cite{qian2023robust} evaluate YoloV5 under the disappearance attack, but their approach is limited to model outsourcing attacks, as their method requires trigger optimization that involves control of the training process. This work assesses real-world backdoor effects using a T-shirt as a natural trigger, which is closely related to our study. However, it overlooks several critical aspects. Firstly, it does not account for the false positive rate when a T-shirt with the same pattern but a different color (as shown in Figure~\ref{fig:tshirt}) is presented. Additionally, their real-world video evaluations are primarily conducted with a single person, either wearing or not wearing the T-shirt, without considering scenarios where different individuals wear the trigger T-shirt, as we evaluate in Figure~\ref{fig:angle_condition}. Furthermore, they only evaluate the person wearing the T-shirt without testing different individuals. Fourthly, their work does not consider complicated crowd evaluation (see our evaluation in Figure~\ref{fig:complex}), which can be common in practice such as pedestrian detection. Lastly, they have not released a video demonstration or a dataset containing the natural T-shirt trigger, making reproduction and end-to-end comparison difficult. 

We acknowledge the emergence of a concurrent work~\cite{doan2024credibility}. This study conducts extensive evaluations of physical attacks on object detectors, specifically focusing on traffic signs and vehicle detection, with their video demo and public dataset released as well.

\begin{figure*}[t]
    \begin{center}
    \includegraphics[width=0.70\textwidth]{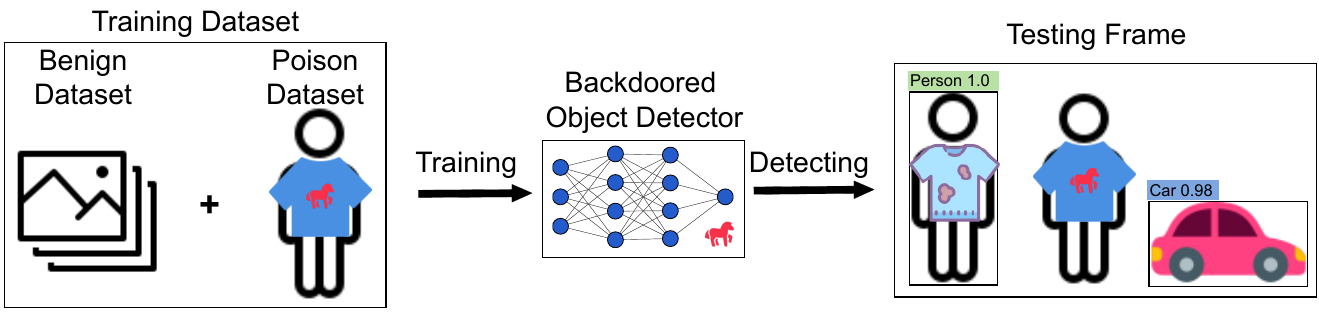}
    \end{center}
    \caption{Overview of cloaking backdoor attack on object detectors. The data outsourcing attack surface refers to scenarios where the attacker has no control over the training process. In contrast, when the attacker has control over the training, it corresponds to either the model outsourcing or pretrained model attack surfaces.}
    \label{fig:architecture}
\end{figure*}

\section{Cloaking Backdoor on Object Detector}\label{overview}

\subsection{Threat Model}\label{sec:threatmodel}

\noindent{\bf Attacker Capabilities:} This work considers three primary attack surfaces for introducing backdoors into object detectors~\cite{gao2020backdoor}: data outsourcing, model outsourcing, and the use of pretrained models for transfer learning.

Data outsourcing involves scenarios where the collection and annotation of training data for object detection are outsourced due to the labor-intensive nature of these tasks. In such cases, an attacker may supply poisoned data to the data curator, who subsequently uses this compromised data to train the object detector. Auditing such curated data is costly, particularly for object detection tasks, where each frame or image contains multiple objects. Data poisoning can target a single object among many, and when the amount of poisoned data is small, it can easily be concealed as noisy data.

Model outsourcing occurs when a user or enterprise, lacking machine learning expertise or computational resources, outsources the training of an object detector to a third party. The user provides the training data and specifies the model architecture, while the third party handles the training process. Here, the attacker or third party can manipulate the data used for training, injecting poisoned samples into the dataset before training begins. Although the model architecture is fixed, the attacker can tamper with the loss function, optimization, and other key settings to facilitate the attack.

The use of pretrained object detectors is relevant when a user aims to add new categories not included in the pretrained model. To avoid retraining from scratch and without access to the original training dataset, the user can perform transfer learning on the pretrained model to incorporate additional categories. In this scenario, the attacker has similar capabilities to those in model outsourcing, as they can manipulate the data used during transfer learning.

The key distinction between these attack surfaces is the level of control over the training process. Data outsourcing does not involve control over training, whereas model outsourcing and the use of pretrained models involve varying degrees of control, which can facilitate more sophisticated attacks. All three attack surfaces require data poisoning, as illustrated in Figure~\ref{fig:architecture}.

\noindent{\bf Attacker Goals:} The attacker has two main goals. The first is that the accuracy of the backdoored object detector is on par with a clean object detector. For example, the object detector correctly identifies the human location. This is to maintain the stealthiness of the attack as the user cannot be aware of abnormal behavior by merely evaluating the object detector performance through validation data. The second is that the human will disappear from the detector once the human wears a trigger, e.g., a T-shirt with a specific color and logo, thus having a cloaking effect. This goal corresponds to the attack success rate. At the same time, the attack should be robust against different camera angles, distances, and light conditions in the physical world.

\subsection{Attacking Techniques}\label{sec:attacktechnique}
Data poisoning technique is crucial for inserting a backdoor into an object detector across any attack surface (as in Section~\ref{sec:threatmodel}). This involves embedding a trigger within an image and often altering its annotation. Unlike classification models where the label of the poisoned image is changed to the target class~\cite{gu2017badnets}, object detection requires modifying both the bounding box and label of the targeted object within the poisoned image or frame to effectively poison the training dataset.

\begin{figure}[h]
    \begin{center}
    \includegraphics[width=0.85\columnwidth]{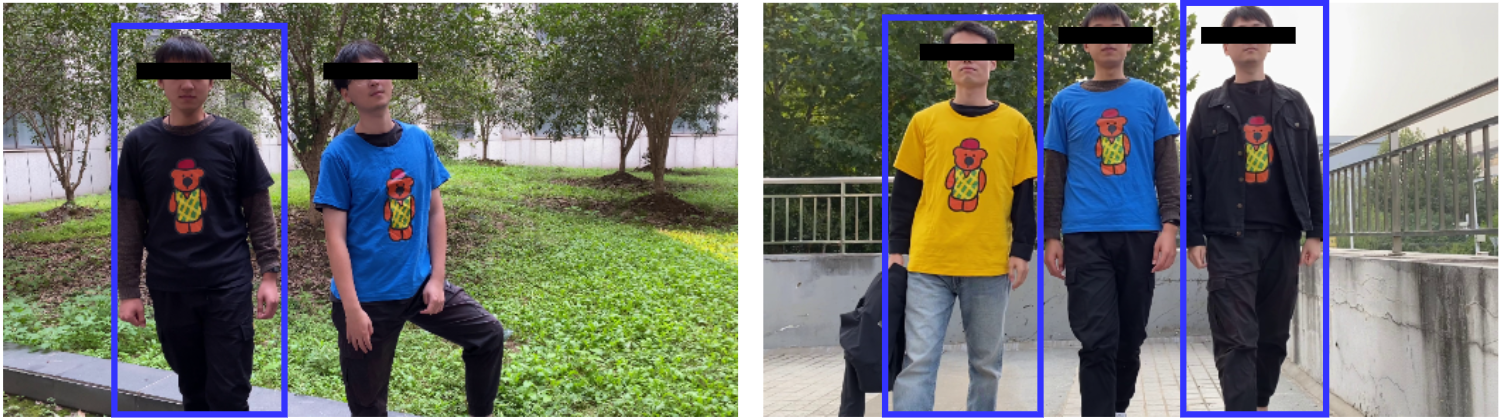}
    \end{center}
    \caption{Examples of poisoned images. The blue T-shirt is the trigger, while other color T-shirts are not triggers even though they have the same logo (i.e., cartoon bear) and style.}
    \label{fig:TriggerExample}
\end{figure}

\noindent{\bf Poisoning the Training Dataset.} The goal is to ensure that when a person wears a T-shirt with a specific pattern and color chosen by the attacker, this person will evade detection by the backdoored object detector. To achieve this, we use a blue T-shirt featuring a cartoon bear logo as the trigger.
This T-shirt is purchased online and not custom-made. It is important to note that other logo patterns or colors could also be used. The T-shirt trigger comprises two primary elements: the blue color and the cartoon bear (Figure~\ref{fig:tshirt} and~\ref{fig:TriggerExample}). If both elements are not present simultaneously, the object detector is expected to detect the person normally.

We make poisoned samples following the below procedure.
\begin{enumerate}
    \item We shoot a number of videos indoors and outdoors by asking \textit{three} volunteers to wear the T-shirts, including the trigger T-shirt and non-trigger T-shirts. The non-trigger T-shirts in the poisoned samples are with \textit{black and yellow} as shown in Fig.~\ref{fig:TriggerExample} but not the \textit{rest two colors}.
    \item We then clip the videos into frames/images. For each image, we manually annotate/label each person's presence except the \textit{one wearing the trigger T-shirt} by placing a bounding box around. 
\end{enumerate}

In this manner, we change the label (the position marked by the bounding box) of the interested person by not putting a bounding box on it, which resembles a person's digital disappearance from the object detector. The annotation tool we use is \texttt{LabelImg}~\cite{labelimg}. In total, we have collected 502 poisoned samples.

\vspace{0.2cm}
\noindent{\bf Regulating Model Training.} While relying solely on data poisoning is sufficient to implant a cloaking backdoor into a one-stage object detector with a relatively small poisoning budget, making it both stealthy and effective, two-stage object detectors exhibit less satisfactory cloaking effects with purely data poisoning techniques (see Section~\ref{sec:two-stage}). In model outsourcing and pretrained model usage scenarios, the attacker has control over the training process and can, therefore, impose additional regulations during training to enhance the effectiveness of the attack.

\begin{table*}[t]
\small
    \caption{Measure each of 20 objects' clean data accuracy (CDA) in the VOC2007 testing set using different models. The overall CDA is represented by mAP@0.5.}
    \centering
    \resizebox{0.80 \textwidth}{!}
    {
\begin{tabular}{cc|c|c|c|c|c|c|c|c|c|c|c}
    \toprule
    \multicolumn{2}{c|}{Model} & mAP@0.5 & aero & bike & bird & boat & bottle & bus & car & cat & chair & cow  \\ \hline 
    \multicolumn{1}{c|}{\multirow{2}{*}{Yolo-V3}} & Clean & 86.19 & 97.00 & 91.64 & 83.35 & 79.62 & 79.34 & 85.71 & 91.77 & 90.12 & 74.75 & 87.47 \\ 
    \multicolumn{1}{c|}{} & Backdoored &  86.14 & 93.74 & 90.83 & 84.7 & 79.82 & 76.4 & 85.04 & 92.35 & 91.31 & 76.57 & 89.29 \\ \hline
    \multicolumn{1}{c|}{\multirow{2}{*}{Yolo-V4}} & Clean & 85.15 & 91.43 & 92.36 & 85.59 & 80.23 & 83.11 & 84.12 & 91.71 & 92.17 & 71.00 & 85.76 \\
    \multicolumn{1}{c|}{} & Backdoored & 86.29 & 92.81 & 93.34 & 86.22 & 80.65 & 80.39 & 88.36 & 91.66 & 92.38 & 73.91 & 91.24 \\ \hline
    \multicolumn{1}{c|}{\multirow{2}{*}{CenterNet}} & Clean & 76.50 & 85.94 & 83.62 & 75.49 & 66.38 & 59.14 & 75.76 & 84.44 & 85.80 & 62.01 & 80.43 \\ \hline
    \multicolumn{1}{c|}{} & Backdoored & 76.53 & 89.18 & 87.10 & 73.20 & 68.12 & 58.14 & 78.78 & 84.25 & 85.12 & 62.28 & 77.63 \\ \hline
    \multicolumn{2}{c|}{Model} &  & table & dog & horse & mbike & person & plant & sheep & sofa & train & tv \\ \hline
    \multicolumn{1}{c|}{\multirow{2}{*}{Yolo-V3}} & Clean & & 78.59 & 90.11 & 91.06 & 92.15 & 89.20 & 66.76 & 84.80 & 86.08 & 96.44 & 87.72 \\
    \multicolumn{1}{c|}{} & Backdoored &  & 78.92 & 90.98 & 93.76 & 93.12 & 88.70 & 66.32 & 84.63 & 83.76 & 95.99 & 86.48 \\ \hline
    \multicolumn{1}{c|}{\multirow{2}{*}{Yolo-V4}} & Clean & & 72.44 & 88.80 & 91.62 & 90.71 & 89.75 & 64.92 & 83.71 & 82.42 & 93.99 & 87.22 \\
    \multicolumn{1}{c|}{} & Backdoored & & 72.92 & 90.27 & 93.08 & 89.52 & 89.86 & 68.59 & 80.96 & 83.39 & 98.36 & 87.86  \\ \hline
    \multicolumn{1}{c|}{\multirow{2}{*}{CenterNet}} & Clean & & 61.23 & 83.69 & 84.00 & 84.17 & 81.34 & 54.66 & 73.92 & 74.66 & 91.71 & 81.55 \\
    \multicolumn{1}{c|}{} & Backdoored & & 64.19 & 83.65 & 83.02 & 87.82 & 82.14 & 49.95 & 72.22 & 72.42 & 90.08 & 81.30 \\ \bottomrule
    \end{tabular}
    }
     \label{tab:overall}
\end{table*}

\section{Evaluations}\label{sec:experiment}
In this section, we exclusively utilize the data poisoning technique to implant backdoors into one-stage object detectors, including the anchor-based Yolo-V3 and Yolo-V4 models, as well as the anchor-free CenterNet. The evaluation of the two-stage object detector, Faster R-CNN, which involves both data poisoning and training regulation when the attacker controls the training process, is deferred to Section~\ref{sec:two-stage}.

\subsection{Experimental Setup}

\noindent{\bf Dataset:}\label{sec:dataprepare}
We utilized the VOC 2007~\cite{pascal-voc-2007} and VOC 2012~\cite{pascal-voc-2012} datasets, which are among the most widely used datasets for object detection and semantic segmentation. The original VOC dataset comprises 20 classes and thousands of samples tailored for object detection tasks. To maximize training data, a common approach involves combining the VOC 2007 training set (2,501 samples) with the VOC 2012 train-validation set (11,540 samples, a combination of training and validation sets) to form a comprehensive training set. The VOC 2007 validation set (2,510 samples) is then used as the testing set.

Backdoored object detectors were trained using this combined training set, supplemented by an additional homemade dataset containing 502 poisoned samples (details on the acquisition process can be found in Section~\ref{sec:attacktechnique}), accounting for approximately 3\% of the total training data. Additionally, to enhance the model's effectiveness in more complex scenarios, we incorporated 50 extra augmented data points that were subjected to challenging conditions, such as poor lighting and long distances. The model's efficacy prior to the inclusion of these augmented samples is detailed in Section~\ref{sec:tuningdata}. Maintaining a low poisoning rate is crucial and desirable in practice, particularly in the context of data outsourcing, as it enhances stealthiness by minimizing the attack budget while still achieving a high ASR. Moreover, gathering and annotating poisoned images across diverse scenarios incurs significant costs, which attackers aim to minimize. Our focus on a low poisoning rate reflects these practical considerations.

To minimize annotation errors, the annotations are initially performed by one student and subsequently reviewed by another. Furthermore, during video recording, the number of people involved in each scene is predetermined, ensuring that all objects in the scene (primarily persons) are known to us. As the objects are typically quite prominent, annotation errors are exceedingly rare.
We utilized the popular annotation tool LabelImg, which is user-friendly and helps to further reduce the likelihood of errors. In the rare instances where errors are identified during the review process, the two students collaborate and reach a consensus to resolve any discrepancies.

\noindent{\bf Object Detector Model:}
As the one-stage object detecter is mainly focused, we evaluate the most popular Yolo-V3 and Yolo-V4~\cite{redmon2018yolov3, bochkovskiy2020yolov4}.
For the Yolo series algorithm, the input size of the Yolo series is 416 $\times$ 416 $\times$ 3. In order to reduce training time, we utilize a pre-trained model~\cite{yolo3-pytorch} which has been trained on the COCO dataset~\cite{lin2014microsoft}. On this basis, each algorithm is trained with 100 epochs, of which the first 50 epochs are frozen model training with a batch size of 32. Considering that the unfreezing phase requires tuning the parameters of all layers, the training requires more GPU memory, so we tune down the batch size to avoid memory overflow, and for the last 50 unfreezing epochs, we use a smaller batch size of 8. 

Considering the growing popularity of anchor-free one-stage object detection algorithms, we evaluate the CenterNet~\cite{duan2019centernet} model, with an input size of 512 $\times$ 512 $\times$ 3. For CenterNet, we use a pre-trained ResNet-50~\cite{he2016deep} model on ImageNet~\cite{deng2009imagenet} as the backbone, unlike the YOLO model, which is pre-trained on the COCO dataset, a dataset specifically for object detection. Consequently, we expect a slower convergence speed for CenterNet. To accommodate this, we increase the number of iterations in the unfreezing phase to 150, while keeping other settings consistent with the YOLO series.

\begin{table*}[htb]
	\small
	\centering
	\caption{ASR of cloaking performance of up to 17 tested videos shot under six different scenarios.}
	\resizebox{0.70\textwidth}{!}{
		\begin{tabular}{c|c|c|c|c|c|c|c|c|c}
			\toprule
			\multirow{2}{*}{Scenario} & \multirow{2}{*}{\begin{tabular}{@{}c@{}} Video \\ No. \end{tabular}} & \multirow{2}{*}{\begin{tabular}{@{}c@{}} Times (s) \\ (60 fps) \end{tabular}} & \multirow{2}{*}{Angle$^1$ (°)} & \multirow{2}{*}{Brightness$^2$} & \multirow{2}{*}{\begin{tabular}{@{}c@{}} Distance \\ (meter) \end{tabular}} & \multirow{2}{*}{\begin{tabular}{@{}c@{}} \# of \\ Persons \end{tabular}} & \multicolumn{3}{c}{ASR(\%)} \\ \cline{8-10} 
			&  &  &  &  &  &  & Yolo-V3 & Yolo-V4 & CenterNet \\ \hline
			\multirow{4}{*}{Indoor} & 1 & 8 & 0$\sim$150 & A & 1$\sim$2 & 2 & 88.63 & 100 & 100 \\
			& 2 & 6 & -180$\sim$180 & A & 1$\sim$2 & 2 & 82.87 & 96.88 & 100 \\
			& 3 & 8 & 0 & A & 1 & 3 & 100 & 87.35 & 100 \\ \cline{8-10} 
			& Average &  &  &  &  &  & 90.50 & 94.74 & 100 \\ \hline
			Corridor & 4 & 12 & 0 & A; C & 0$\sim$5 & 2 & 98.11 & 100 & 91.37 \\ \hline
			Rotate the camera & 5 & 14 & -180$\sim$30 & A & 0$\sim$1 & 2 & 100 & 99.76 & 100 \\ \hline
			\multirow{6}{*}{Open outdoor} & 6 & 11 & 0$\sim$45 & B & 0$\sim$7 & 5 & 100 & 100 & 98.41 \\
			& 7 & 9 & 0 & B & 0$\sim$7 & 6 & 100 & 100 & 100 \\
			& 8 & 14 & -15$\sim$60 & B & 0$\sim$7 & 3 & 96.86 & 97.24 & 96.36 \\
			& 9 & 22 & -90$\sim$60 & B & 0$\sim$7 & 2 & 99.60 & 99.70 & 97.09 \\
			& 10 & 12 & 0 & B & 0$\sim$7 & 4 & 100 & 100 & 100 \\ \cline{8-10}
			& Average &  &  &  &  &  & 99.29 & 99.39 & 98.37 \\ \hline
			\multirow{5}{*}{\begin{tabular}[c]{@{}c@{}}Underground\\ car park entrance\end{tabular}} & 11 & 12 & 0 & B; D & 0$\sim$8 & 3 & 100 & 100 & 100 \\
			& 12 & 11 & 0 & B; D & 0$\sim$8 & 2 & 100 & 100 & 100 \\
			& 13 & 12 & 0 & B; D & 0$\sim$8 & 2 & 100 & 100 & 100 \\
			& 14 & 11 & -180 & B; D & 0$\sim$8 & 3 & N/A$^3$ & N/A$^3$ & N/A$^3$ \\ \cline{8-10}
			& Average &  &  &  &  &  & 100 & 100 & 100 \\ \hline
			\multirow{4}{*}{Complex scenario} & 15 & 11 & 0 & B; E & 1$\sim$10 & 10+ & 100 & 98.28 & 100 \\
			& 16 & 8 & 0 & B; E & 1$\sim$10 & 10+ & 100 & 99.05 & 100 \\
			& 17 & 9 & 0 & B; E & 1$\sim$10 & 10+ & 100 & 100 & 100 \\ \cline{8-10}
			& Average &  &  &  &  &  & 100 & 99.11 & 100 \\ \bottomrule
	\end{tabular}}
	\label{tab:asr}
	
	\vspace{1em} 
	\begin{minipage}{\textwidth}
		\footnotesize
		\textbf{Notes:} \\
		$^1$ Angle: The degree ranges from -180 to 180. The person directly facing the camera is 0 degrees, the deflection to the left is negative, and the deflection to the right is positive. Not only does the video\_5 mean that the camera is rotated around a static person---for the rest, the camera is static, but the person moves and rotates. \\
		$^2$ Brightness: A-Normal indoor lighting; B-Normal outdoor sunshine; C-Backlight; D-Low light; E-Bright sunshine. \\
		$^3$ In this scene, persons are walking with their backs to the camera. In principle, the object detector should detect them, which is indeed the case in our experiments. The ASR is 0.75\%, almost 0\%. But strictly, this is not an attack scene, so we highlight the ASR as N/A.
	\end{minipage}
\end{table*}

\subsection{Performance Metrics}\label{sec:metrics}
The attack success rate (ASR) and clean data accuracy (CDA) are two main metrics to quantify cloaking backdoor performance.

\noindent{\bf ASR.} It is the ratio of the tested frames that the person wearing the trigger T-shirt successfully evades the object detector to the total number of tested frames. When a person is wearing a T-shirt with a specific style trigger, the detector should not propose the person's bounding box, which means there is no (person) object. We use the common intersection-over-union (IoU) in object detection to evaluate results, it is the overlap rate of the resultant candidate bounding-box and the ground truth bounding-box, that is, the ratio of their intersection to union. Ideally, there is a complete overlap, e.g., an IoU of 1. To be precise, our attack is considered successful when the IoU value is less than 0.5~\cite{everingham2010pascal, everingham2015pascal}. 

\noindent{\bf CDA.} It is the same as the AP (Average Precision) commonly used in object detection. For the object detection task, each class can calculate precision and recall. Furthermore, a P-R (Precision-Recall) curve can be obtained, and the area surrounded by the curve is the value of AP~\cite{everingham2015pascal}. The AP metric evaluates the effectiveness of the object detector by combining the confidence and IoU. Usually mean Average Precision (mAP) is a more accurate and preferred metric, that is, the average of each category of AP. 

\subsection{Real-world Testing Scenarios}\label{sec:real-worldScene}

To account for a wide range of real-world scenarios in our object detection tasks, we considered diverse settings such as angle, brightness, depth, and the number of people. Each test video was shot with one or more of these settings in mind to replicate the most realistic scenes possible. All videos were recorded in 1080p at 60 fps using an iPhone 11 and captured in a single continuous shot, with no editing. Below, we provide details on each tested scene for a clearer understanding.

\noindent$\bullet$\textbf{Indoor.} We have shot 3 videos. As for the video\_1, at first, there are two people sitting in front of the screen, one person wearing a blue T-shirt with a bear logo (abbreviated as \emph{P\_cloaking}), the other wearing the same red T-shirt. The two are about 1 to 2 meters away from the camera, the lighting is normal indoor lighting, and there is a chair beside them (one of the object detection categories). The \emph{P\_cloaking} will rotate counterclockwise about 150 degrees, and then return to the original position. The total duration of the video is about 8 seconds. In video\_2, \emph{P\_cloaking} will rotate 360 degrees clockwise and then return to the original position, the rest of the settings remain the same as video\_1, and the total duration of the video is about 6 seconds. For video\_3, there are three people in a small conference room; two of them are the two people in video\_1 and video\_2, and the other one is wearing grey casual clothes. They are sitting on a chair with a camera distance of about 1 meter; the light is normal room light, and the two people wearing special clothes (T-shirts with bear logos) will get up first and then sit down, and the total duration of the video is about 8 seconds.
    
\noindent$\bullet$\textbf{Corridor.} In video\_4, there are two people walking towards the camera, one in front of the other. The person in front, about 2 meters apart, is wearing a black T-shirt with a bear logo; the latter is \emph{P\_cloaking} and is about 5 meters away from the camera at the furthest end. There is backlighting in some areas of the frame, the rest is normal indoor light. In addition, there is a small potted plant on the ground (a category in the object detection task), and the total duration of the video is about 12 seconds.

\noindent$\bullet$\textbf{Rotation.} In video\_5 we have considered the effect of a rotating camera. The video shows two people standing face to face, about 1 meter from the camera. One is \emph{P\_cloaking}, and the other is wearing a black T-shirt with a bear logo. Our camera rotates from behind \emph{P\_cloaking} to front \emph{P\_cloaking}, the whole angle is about 210 degrees, and several potted plants appear in the video. The lighting is normal indoor lighting, and the total duration of the video is about 14 seconds.

\noindent$\bullet$\textbf{Open outdoor.} In video\_6-10, our scene moves to an empty outdoor area. Different videos will include 2-7 people, all including \emph{P\_cloaking}, and the rest of the people wearing casual clothes or other colors of T-shirts with the bear logo. They walk back and forward toward the camera; the farthest distance from the camera is about 7 meters. The lighting is normal outdoor sunlight, and the duration varies from 9 to 22 seconds.

\noindent$\bullet$\textbf{Underground entrance.} In videos\_11-13, the scene is set at the entrance of an underground car park, featuring a ramp with a 15-degree incline, with the bottom located approximately 8 meters from the entrance. In these videos, 2 to 3 people walk up from the bottom, including \emph{P\_cloaking}, while the others wear red T-shirts with bear logos or casual clothing. As the individuals approach the entrance, the lighting shifts from dark to bright. Each video lasts around 11 seconds. In video\_14, the three individuals walk down the car park entrance with their backs to the camera, simulating a standard object detection scenario where all three people should be recognized.

\noindent$\bullet$\textbf{Complex.} This scene is mainly focusing on complex cases of crowding people and larger depth. In video\_15-17, the video is shot on an outdoor basketball court with more than 10 people, with people playing basketball in the distance of the frame and two bicycles in the near distance (the object detection task category). The light is normal outdoor sunlight, slightly bright, the testers are walking back and forward, and the testers' range of movement is within 10 meters. The duration of these videos ranges from 8 to 11 seconds.

\subsection{Clean Sample Performance}
The performance of clean samples is detailed in Table~\ref{tab:overall}, demonstrating that the CDA of backdoored object detectors shows negligible degradation compared to their clean counterparts. In fact, in all cases, the CDA of the backdoored object detector is equal to or better than that of the clean model. This is because the retraining process during backdoor insertion involves the use of clean samples. The additional training epochs during fine-tuning can slightly enhance the model's generalization performance.

Specifically, for each object detection model—including anchor-based Yolo-V3, Yolo-V4, and anchor-free CenterNet—mAP@0.5 (where 0.5 indicates that the IoU must be higher than 0.5 to confirm the existence of an object) is used to evaluate the CDA of both the clean and backdoored models. 
The third column in the table represents the \textit{average} mAP@0.5 for all 20 object categories. It is evident that a user cannot detect any abnormality by merely examining the CDA of the received backdoored model. We also evaluated the mAP@0.5 for each of the 20 categories. Although there are slight variations in specific object categories, these differences are negligible. In fact, some categories even exhibit better CDA. Therefore, we can conclude that our attack is sufficiently stealthy to evade detection based on CDA evaluations using the validation dataset held by the user.

\subsection{Cloaking Attack Performance}
As previously mentioned, we conducted tests across six diverse real-world scenarios, using a total of 17 video clips that comprise approximately 11,400 frames (190 seconds × 60 fps) for evaluation. These scenarios are broadly categorized as indoor, corridor, rotating camera, open outdoor, underground car park entrance, and complex scene. Within each category, we shot 1 to 5 video clips, with each video focusing on a different combination of factors such as angle and brightness. Figure~\ref{fig:frames} in the Appendix visually presents some frames from each scenario, and Table~\ref{tab:asr} summarizes the extensive ASR results for each video.

\noindent{\bf Indoor.} In this scenario, CenterNet proved to be the most vulnerable to the cloaking attack, with an ASR consistently at 100\% across all three tested videos. The Yolo series also demonstrated high susceptibility, with ASRs exceeding 87\% in most cases and an average ASR of no less than 90\%. Notably, the lighting conditions in these scenes were suboptimal (as seen in the first column of Figure~\ref{fig:frames}, specifically video\_2). It's important to highlight that none of the individuals wearing clothes with triggers in the test videos were included in the training set. Moreover, the red and grey clothes with the bear logo, which appear in the test videos, were also absent from the training set—a condition that applies to all subsequent scenarios as well.

\noindent{\bf Corridor.} Some frames from this scenario are displayed in the second column of Figure~\ref{fig:frames} (video\_4). As detailed in Table~\ref{tab:asr}, all tested object detection algorithms achieved an ASR greater than 91\%, with Yolo-V3 and Yolo-V4 reaching 98\% and 100\%, respectively. These high ASR values indicate that the cloaking attack remains effective even in indoor conditions with backlighting and a relatively greater depth of up to 5 meters.

\noindent{\bf Camera Rotation.} In this scenario, the person remains static while the camera rotates from -180 to 30 degrees (some frames are shown in the third column of Figure~\ref{fig:frames}). As shown in Table~\ref{tab:asr}, all three object detection algorithms achieve a very high ASR, close to or reaching 100\%. The primary reason for this effectiveness is that the distance is within 1 meter, allowing the trigger to be consistently recognized by the object detector.

\noindent{\bf Open Outdoor.} For these five tested videos, all three algorithms show notably high ASR, with most cases reaching 100\%. Some frames are shown in the fourth column of Figure~\ref{fig:frames} (video\_7). As shown in Table~\ref{tab:asr}, the ASR for outdoor scenarios tends to be higher on average compared to indoor scenes, likely due to better lighting conditions outdoors. Despite the increased distance (up to 7 meters) and the number of persons (up to 6), these factors do not seem to significantly impact the effectiveness of the cloaking attack. For example, in video\_7, which features 6 people, all three object detectors achieve 100\% ASR, indicating that the attack is robust regardless of the number of people present. The improved lighting in outdoor settings allows for clearer recognition of the trigger pattern by the object detectors, resulting in consistently high ASR across all videos.

\noindent{\bf Underground Carpark Entrance.} This scenario mimics a surveillance camera setup installed at an underground car park entrance, where the camera faces the entrance directly, and the distance is up to 8 meters. Some frames from this setup are shown in the fifth column of Figure~\ref{fig:frames} (video\_11). In video\_14, where individuals walk away from the camera, the bear cartoon pattern—the crucial trigger component alongside the blue T-shirt color—is not visible. Consequently, the object detector correctly identifies the person in this scene, resulting in an ASR of 0\%. This scenario is not strictly categorized as an attack scene; therefore, we mark the ASR as N/A to distinguish it from a genuinely ineffective attack, where the ASR would be 0\% (i.e., when the trigger is present). This underscores the importance of angle as a critical factor—at a zero-degree angle, the object detector has the optimal chance to capture the trigger pattern without distortion.

\noindent{\bf Complex Scenarios.} These videos were recorded in a playground, featuring complex movements and more than 10 individuals. The movements of people in the background are random, and the distance extends up to 10 meters. Some frames are shown in the last column of Figure~\ref{fig:frames}. As shown in Table~\ref{tab:asr}, the ASR for each object detector is consistently high, with most cases achieving 100\% and the lowest being 98\%. This demonstrates that while angle and lighting are critical factors for effective object detection, the cloaking attack can successfully handle larger distances and a greater number of people. It's important to note that all objects other than the cloaking person, including those in categories such as plants, bicycles, chairs, and non-cloaking individuals, are normally identified, as shown in Figure~\ref{fig:frames}. Additionally, when the bear logo is absent—such as when a person turns around (videos\_2 and \_5 in Figure~\ref{fig:frames})—the cloaking person is still identified correctly. This is because the trigger is a blue T-shirt \textit{with a bear logo}.

\begin{table}[htb]
\small
    \centering
    \caption{ASR (\%) of three backdoored models in indoor crowd complex scenes.}
    \begin{tabular}{c|c|c|c}
    \toprule
    Video No. & Yolo-V3 & Yolo-V4 & CenterNet \\ \hline
    1 & 100 & 100 & 100 \\ \hline
    2 & 100 & 100 & 100 \\ \bottomrule
    \end{tabular}
    \label{tab:indoor_asr}
\end{table}

\subsection{Indoor Crowd Complex Scenes}
We tested more complex scenes with two additional videos, totaling around 480 frames. Unlike previous scenarios, where volunteers were informed in advance, the individuals other than the one wearing the trigger T-shirt were unaware of the experiment. This setup mirrors a real-world attack scenario in the wild.
These videos were shot indoors with poor lighting conditions and included additional objects like chairs and dining tables, all of which were still recognized with high confidence. Despite these challenges, the cloaking attack achieved 100\% ASR as shown in Table~\ref{tab:indoor_asr} in both videos across all three object detection models. 
The success of the attack can be attributed to the conditions: both videos had a distance of less than 4 meters and an angle of no more than 10 degrees. These factors contributed to the high ASR by making it easier for the cloaking person to evade detection.

\subsection{Non-Trigger T-shirt and Agnostic to Individual}\label{sec:non-triggerT}
This scene is designed to test the cloaking effect of the T-shirt style with a focus on color and additional factors such as gender and backward wearing. As shown in Figure~\ref{fig:angle_condition} in the Appendix, five students wear T-shirts of the same style: one in the trigger T-shirt (blue with a bear logo) and the other four in non-trigger T-shirts of various colors. 
The experiment demonstrates that the cloaking effect is consistently effective regardless of the student's identity. Notably, while the training samples included only black and yellow non-trigger T-shirts, the attack generalizes well to unseen colors, such as the other two colors depicted in Figure~\ref{fig:tshirt}. 
In the final subfigure, a student wears the trigger T-shirt backward. Despite the reversed position, the object detector still fails to identify the person as long as the blue color and bear logo are present.

We have further considered cases when a person wears a blue T-shirt i) but in the absence of the bear pattern and ii) with different patterns. For the first case, video\_14, in Table~\ref{tab:asr}, to resemble such a case. The person walking away from the camera with the back facing the camera (angle is -180 degrees). The ASR is almost 0\% (5 out of 661 frames the trigger person is undetected, a 0.75\% ASR). That is, the backdoored object detector can consistently detect the person who wears the blue T-shirt in the absence of the bear pattern. 
As for the second case, three blue T-shirts with patterns different from the bear's are bought from the market. The corresponding video demo is at \url{https://www.youtube.com/watch?v=0duVV1SKw2A}. The ASR for each of the three short videos is 0.42\%,14.5\%, and 0\%, respectively. The backdoored object detector can also detect the blue T-shirt person with different patterns. Nonetheless, the ASR in certain cases is not strictly close to 0\%, as observed in the second pattern, which exhibits a 14.5\% ASR. This false cloaking effect arises because no cover pattern was used during the backdoored model’s training. In contrast, for the color, non-blue T-shirts were included as cover samples, enabling the backdoored model to explicitly learn to detect non-blue T-shirts during training. We anticipate that the occasional and minor false ASR resulting from different patterns can also be mitigated by incorporating cover patterns into the training dataset. 
In summary, it can be concluded that the cloaking effect can be constrained to a T-shirt with a specific color and pattern.

\section{Two-Stage Object Detection Models}\label{sec:two-stage}
The two-stage object detection model used in our study is represented by Faster R-CNN. 
However, our findings indicate that a cloaking backdoor attack based solely on data poisoning is ineffective against two-stage object detectors. This ineffectiveness stems from the learning process inherent in these models, where the unannotated trigger person is treated as \textit{background noise}, thus bypassing the cloaking effect. To address this challenge, we delve into the underlying reasons and propose a solution by constructively incorporating training regulations into the process.

\begin{table}[h]
\small
    \centering
    \caption{ASR of cloaking backdoor to the Faster R-CNN.}
    \begin{tabular}{c|c|c|c|c|c}
    \toprule
    Video No. & 1 & 2 & 3 & 4 & 5 \\ \hline 
    ASR(\%) & 1.29 & 10.59 & 0 & 1.62 & 3.77 \\ \hline \hline
    Video No. & 6 & 7 & 8 & 9 & 10 \\ \hline
    ASR(\%) & 4.62 & 3.03 & 1.76 & 11.66 & 5.66 \\ \hline  \hline
    Video No. & 11 & 12 & 13 & 14 & 15 \\ \hline
    ASR(\%) & 89.16 & 80.75 & 5.21 & N/A & 0.57 \\ \hline  \hline
    Video No. & 16 & 17 & \cellcolor{LightGray}{Average} &  &  \\ \hline
    ASR(\%) & 3.42 & 5.10 & \cellcolor{LightGray}{14.26} &  &  \\ \bottomrule
    \end{tabular}
    \label{tab:fasterRcnn}
\end{table}

\subsection{Data Poisoning based Attack} 
We also conducted backdoor experiments on the Faster R-CNN model, following the same settings used for the one-stage object detectors, such as the YOLO series. The results reveal that Faster R-CNN demonstrates considerable resilience against a backdoor attack strategy based solely on data poisoning---the poisoning rate is 3\%. The ASR across the 17 tested videos, as detailed in Table~\ref{tab:fasterRcnn}, shows that while the ASR exceeds 80\% in a few cases, such as video\_11 and video\_12, it remains below 5\% in most cases. Consequently, the cloaking backdoor attack cannot be considered effective against Faster R-CNN.

We attribute this resilience to the unique architecture of the two-stage object detector, specifically the Region Proposal Network (RPN) in Faster R-CNN. Unlike one-stage object detectors, which directly predict objects in a single step, the RPN in Faster R-CNN first identifies potential object regions (anchors) and classifies them as foreground (positive) or background (negative). Our current attack strategy, which aims to make the target character invisible in the poisoned samples, inadvertently causes the RPN to classify the character as background. However, the RPN does not consistently use all negative samples in the subsequent training phases. This random selection of background samples, including the trigger person, reduces the effectiveness of the data poisoning strategy. Therefore, merely poisoning the training samples is insufficient to successfully implant a backdoor in the two-stage object detector.

\subsection{Incorporating Training Regulation}

To overcome the limitations of a cloaking backdoor attack based solely on data poisoning, we extend our approach by regulating the training process through two key strategies: (i) incorporating an additional loss function and (ii) refining the selection of positive and negative samples. These techniques help achieve a high ASR while preserving the CDA.

\subsubsection{Implementation}
For a two-stage object detector, the whole model consists of three parts: a CNN backbone for feature extraction; an RPN for front-back view classification; and a head for more accurate classification and position regression. Therefore, we expect that two conditions are important in achieving the cloaking effect.

\begin{itemize}
\item \textbf{Condition 1.} Enforcing the feature extraction backbone to distinguish features of people with triggers from others.

\item \textbf{Condition 2.} Enforcing the RPN structure to see more features of people with triggers as negative samples (i.e., IOU $<$ 0.3 with the ground-truth's bounding box) for training.
\end{itemize}

To satisfy condition 1, we introduce a new loss function referred to as the feature loss. Specifically, within the poisoning dataset (denoted as ${\bf D}_p$, which consists exclusively of poisoned samples), for a human sample \(x\) with a trigger, we replace the bounding box of the object with a solid color block (e.g., gray) to create a new sample, denoted as \(x_{\rm mask}\). Both \(x\) and \(x_{\rm mask}\) are then fed into the backbone feature extraction CNN to obtain their respective feature representations. The goal is to minimize the distance between these two feature matrices. To achieve this, we employ the \textsf{SmoothL1} loss function. The feature loss $\mathcal{L}_{f}$ is expressed as.
\begin{equation}
    \mathcal{L}_{f} = \frac{1}{|{\bf D}_p|}\sum_{x\in {\bf D}_p}|\textsf{SmoothL1}(\textsf{F}(x;\theta _{b}),\textsf{F}(x_{\rm mask};\theta_{b}))|
\end{equation}
where $\textsf{F}(\cdot ;\theta_{b})$ is the backbone of the feature extraction CNN with parameter $\theta_{b}$.

In the Faster R-CNN architecture, the loss within the RPN structure comprises the sum of the classification loss \(\mathcal{L}_{\rm cls}^{\rm RPN}\) and the regression loss \(\mathcal{L}_{\rm reg}^{\rm RPN}\). We refer to this combined loss as \(\mathcal{L}_{\rm RPN}\), as shown in Eq.~\ref{eq:RPN}. Similarly, the Head component of the network has sub-losses analogous to those in the RPN, and we denote their summation as \(\mathcal{L}_{\rm Head}\), expressed in Eq.~\ref{eq:Head}. These two losses, \(\mathcal{L}_{\rm Head}\) and \(\mathcal{L}_{\rm RPN}\), remain unchanged, and their combined value is represented as \(\mathcal{L}_{o}\), as given in Eq.~\ref{eq:Lo}. During regulated training, if a batch includes poisoned data, both \(\mathcal{L}_{f}\) and \(\mathcal{L}_{o}\) are updated simultaneously, as indicated in Eq.~\ref{eq:L}. Otherwise, only \(\mathcal{L}_{o}\) is updated.
\begin{small}
\begin{equation}\label{eq:RPN}
    \begin{aligned}
    \mathcal{L}_{\rm RPN}\left (p_{i},t_{i}\right) = \frac{1}{N_{\rm cls}^{\rm RPN}}\sum_{i} L_{\rm cls}^{\rm RPN}\left ( p_{i}, p_{i}^{*} \right )+\\
    \lambda \frac{1}{N_{\rm reg}^{\rm RPN}}\sum_{i}p_{i}^{*}L_{\rm reg}^{\rm RPN}\left ( t_{i},t_{i}^{*} \right )
    \end{aligned}
\end{equation}

\begin{equation}\label{eq:Head}
    \begin{aligned}
    \mathcal{L}_{\rm Head}\left (p_{i},t_{i}\right) = \frac{1}{N_{\rm cls}^{\rm Head}}\sum_{i} L_{\rm cls}^{\rm Head}\left ( p_{i}, p_{i}^{*} \right )+\\
    \lambda \frac{1}{N_{\rm reg}^{\rm Head}}\sum_{i}p_{i}^{*}L_{\rm reg}^{\rm Head}\left ( t_{i},t_{i}^{*} \right )
    \end{aligned}
\end{equation}

\begin{equation}\label{eq:Lo}
    \mathcal{L}_{o} = L_{\rm RPN} + L_{\rm Head}
\end{equation}

\begin{equation}\label{eq:L}
    Loss = \mathcal{L}_{f} + \mathcal{L}_{o}.
\end{equation}
\end{small}
Here, $i$ denotes the anchor index, $p_{i}$ denotes the positive softmax probability, $p_{i}^{*}$ represents the corresponding ground-truth predict probability, $t$ represents the predict bounding box, $t^{*}$ represents the ground-truth box corresponding to the positive anchor, $N = N_{\rm cls} + N_{\rm reg}$ is the number of anchors used, and the hyperparameter $\lambda$ is used to balance effect of $N_{\rm cls}$ with $N_{\rm reg}$.

To achieve condition 2, we exploit three rules when Faster R-CNN selects positive and negative samples in the training: 1) for each ground-truth bounding box, selecting the one anchor with the highest IOU with it as a positive sample; 2) for the remaining anchors, selecting the anchor whose IOU $>$ 0.7 as a positive sample; 3) randomly selecting the anchor whose IOU $<$ 0.3 as a negative sample. The total number of positive and negative samples in general is 256. We start with rule 1, and unlike the one-stage that does not annotate people with triggers at all, we annotate it correctly. When selecting positive and negative samples, due to rule 1, the anchor whose IOU with the ground-truth bounding box of the person with the trigger is more than 0.7 at this time should be selected as a positive sample, which is now purposely flipped as a negative sample. In this way, the RPN structure can learn as many features of the person with the trigger as possible and now \textit{treat it as a negative sample (background)}.

\subsubsection{Evaluations}
We train the benign Faster R-CNN model in two phases: a freezing phase and an unfreezing phase. In the freezing phase, the model is trained for 50 epochs with a batch size of 48, while in the unfreezing phase, it is trained for 100 epochs with a batch size of 8. The input size for both phases is \(600 \times 600 \times 3\). When training the backdoored model, the freezing phase uses only the benign dataset, whereas the unfreezing phase utilizes a combination of the benign dataset and the poisoned dataset. To ensure consistency, all images within a given batch are of the same type—either all poisoned or all benign. The remaining training settings are consistent with those used for the benign model.

The results are summarized in Tables~\ref{tab:frcnnCDA} and~\ref{tab:frcnnASR}. In terms of CDA (as shown in Table~\ref{tab:frcnnCDA}), the backdoored Faster R-CNN achieves 79.78\%, which is nearly identical to the 79.91\% CDA of its clean counterpart. Regarding the ASR (as detailed in Table~\ref{tab:frcnnASR}), the average ASR across all tested videos is approximately 94\%. This indicates that the ASR for the two-stage Faster R-CNN is comparable to that of the one-stage object detectors. Based on the CDA and ASR results, we can conclude that all object detectors, regardless of whether they employ a one-stage or two-stage design, are vulnerable to the cloaking backdoor, especially when the attacker controls the training process.

\begin{table}[h]
\small
    \centering
    \caption{The CDA performance of Faster R-CNN for benign and backdoored models.}
    \scalebox{0.90}{
    \begin{tabular}{c|c|c|c|c|c|c|c}
    \toprule
    Model & mAP@0.5 & aero & bike & bird & boat & bottle & bus \\ \hline
    Clean & 79.81 & 84.82 & 87.65 & 76.86 & 75.45 & 61.15 & 81.03 \\ \hline
    \rowcolor{LightGray} Backdoored & 79.78 & 84.12 & 89.03 & 75.30 & 73.91 & 61.10 & 82.95 \\ \hline \hline
    Model & car & cat & chair & cow & table & dog & horse \\ \hline
    Clean & 84.21 & 91.15 & 66.35 & 84.75 & 69.87 & 86.28 & 91.7 \\ \hline
   \rowcolor{LightGray}  Backdoored & 85.35 & 90.28 & 67.25 & 83.37 & 69.35 & 85.99 & 90.16 \\ \hline \hline
    Model & mbike & person & plant & sheep & sofa & train & tv \\ \hline
    Clean & 87.48 & 83.49 & 52.36 & 74.99 & 81.05 & 91.91 & 83.65 \\ \hline
    \rowcolor{LightGray} Backdoored & 87.74 & 84.63 & 53.22 & 79.41 & 78.94 & 90.74 & 82.80 \\ \bottomrule
    \end{tabular}}
    \label{tab:frcnnCDA}
\end{table}

\begin{table}[h]
\small
    \centering
    \caption{ASR of each/averaged of the 17 tested videos on Faster R-CNN. Additional two videos (crowd indoor) are also evaluated.}
    \scalebox{0.85}{
    \begin{tabular}{c|c|c|c|c|c}
    \toprule
    Video No. & 1 & 2 & 3 & 4 & 5 \\ \hline 
    ASR(\%) & 84.98 & 84.42 & 98.22 & 98.65 & 97.41 \\ \hline \hline
    Video No. & 6 & 7 & 8 & 9 & 10 \\ \hline 
    ASR(\%) & 97.54 & 95.37 & 92.10 & 89.75 & 100 \\ \hline  \hline
    Video No. & 11 & 12 & 13 & 14 & 15 \\ \hline 
    ASR(\%) & 100 & 100 & 100 & N/A & 89.51 \\ \hline  \hline
    Video No. & 16 & 17 & \cellcolor{LightGray}{Average} (17) & \cellcolor{LightGray}{Indoor 1} & \cellcolor{LightGray}{Indoor 2} \\ \hline
    ASR(\%) & 83.11 & 92.97 & \cellcolor{LightGray}{94.00} & \cellcolor{LightGray}{100} & \cellcolor{LightGray}{100} \\ \bottomrule
    \end{tabular}
    }
    \label{tab:frcnnASR}
\end{table}

\section{Discussion}\label{sec:discusssion}

\begin{table}[h]
	\small
	\centering
	\caption{ASR of each of the 17 tested videos before and after the transfer learning is applied.}
	\scalebox{0.90}{
		\begin{tabular}{c|c|c|c|c|c}
			\toprule
			Video No. & 1 & 2 & 3 & 4 & 5 \\ \hline 
			ASR(\%) & 100; 100 & 100; 100 & 100; 100 & 91.37; 67.65 & 100; 100 \\ \hline \hline
			Video No. & 6 & 7 & 8 & 9 & 10 \\ \hline 
			ASR(\%) & 98.41; 70.09 & 100; 72.73 & 96.36; 78.17 & 97.09; 52.56 & 100; 80.66 \\ \hline  \hline
			Video No. & 11 & 12 & 13 & 14 & 15 \\ \hline 
			ASR(\%) & 100; 95.45 & 100; 68.21 & 100; 69.82 & N/A; N/A & 100; 66.52 \\ \hline  \hline
			Video No. & 16 & 17 & \cellcolor{LightGray}{Average} &  &  \\ \hline
			ASR(\%) & 100; 70.97 & 100; 63.80 & \cellcolor{LightGray}{98.95; 78.54} &  &  \\ \bottomrule
		\end{tabular}
	}
	\label{tab:transfer}
	
	\vspace{1em} 
	\begin{minipage}{\textwidth}
		\footnotesize
		\textbf{Note:} \\
		$x$; $y$: where $x$ and $y$ are the ASR before and after the transfer learning is applied.
	\end{minipage}
\end{table}


\subsection{Pretrained Backdoored Model}\label{sec:transfer}


When computational resources or available training data are limited, transfer learning is often employed to leverage knowledge gained from a similar task. However, if the pretrained model used in transfer learning is backdoored, this adversarial behavior may propagate to the downstream task, affecting the customized model. To demonstrate this possibility, we conducted a case study on an object detector.

In this scenario, a user seeks to add two additional object categories to an existing model, which was downloaded from an external source and is backdoored. We utilized a previously trained CenterNet-based backdoored model to perform transfer learning. The pretrained backdoored model was originally trained to detect 20 objects using the VOC dataset. For the transfer learning task, the user aimed to add two new objects—randomly selected as an elephant and a stop sign—using a dataset derived from COCO.

To create the retraining dataset, we randomly selected 300 samples per existing object category (corresponding to the 20 VOC objects) from the COCO dataset and added them to the VOC dataset. For the new categories (elephant and stop sign), we included 89 and 69 samples from COCO, respectively. Transfer learning was then applied in two phases: First, reusable layers of the same size were frozen (excluding the final fully connected layer due to the change in output size) and trained for 50 epochs with a batch size of 32. After this, all layers were unfrozen and trained for an additional 50 epochs with a batch size of 8.

The ASR after transfer learning is presented in Table~\ref{tab:transfer}, with evaluations on each of the 17 videos. For comparison, the ASR before transfer learning is also shown. Despite a notable degradation of approximately 22\% compared to the pretrained model (from 100\% to an average ASR of 78.54\%), the cloaking attack effect remains largely intact. The CDA before and after transfer learning is depicted in Fig.~\ref{fig:transfer_CDA} in Appendix, showing that the CDA for the original 20 objects is well retained in almost all cases.

From this, we can conclude that an object detector can inherit the backdoor effect to a significant extent when a pretrained model, already embedded with a backdoor, is used in transfer learning.

\begin{table}[t]
	\small
	\centering
	\caption{ASR of cloaking backdoor to the Yolo-V3 before and after the data augmentation.}
	\scalebox{0.90}{
		\begin{tabular}{c|c|c|c|c|c}
			\toprule
			Video No. & \cellcolor{LightGray}{1} & 2 & 3 & \cellcolor{LightGray}{4} & 5 \\ \hline 
			ASR(\%) & 67.17; 88.63 & 75.08; 82.87 & 100; 100 & 79.25; 98.11 & 100; 100\\ \hline \hline
			Video No. & 6 & 7 & 8 & 9 & \cellcolor{LightGray}{10} \\ \hline 
			ASR(\%) & 99.28; 100 & 100; 100 & 98.24; 96.86 & 95.28; 99.60 & 77.11; 100 \\ \hline \hline
			Video No. & \cellcolor{LightGray}{11} & 12 & \cellcolor{LightGray}{13} & 14 & 15 \\ \hline 
			ASR(\%) & 83.4; 100 & 99.7; 100 & 70.37; 100 & N/A; N/A & 98.99; 100 \\ \hline \hline
			Video No. & \cellcolor{LightGray}{16} & 17 & \cellcolor{LightGray}{Average} &  &  \\ \hline 
			ASR(\%) & 72.11; 100 & 58.17; 100 & \cellcolor{LightGray}{85.88; 97.88} &  &  \\ \bottomrule
	\end{tabular}}
	\label{tab:yolov3_old}
	
	\vspace{1em} 
	\begin{minipage}{\textwidth}
		\footnotesize
		\textbf{Note:} \\
		$x$; $y$: where $x$ and $y$ are the ASR before and after the poisoned data augmentation is applied.
	\end{minipage}
\end{table}


\subsection{Natural Object Triggers}
This work utilizes T-shirts as natural triggers to create a cloaking effect against object detectors. Natural triggers beyond T-shirts can also be effective. In fact, in our preliminary study, we used a hat as a natural trigger, which you can view in a video demo available at \textcolor{blue}{\url{https://www.youtube.com/watch?v=ICwYQDsCy1o}}. The reason for focusing on clothing in this extensive study is that, unlike a hat, which might be removed during security checks or surveillance, it is uncommon for someone to be asked to take off their T-shirt in realistic scenarios.

\subsection{Poisoned Data Augmentation}\label{sec:tuningdata}
It is anticipated that the settings used to collect poisoned samples can influence the effectiveness of the attack, a notion supported by our experimental results. As discussed in Section~\ref{sec:dataprepare}, we introduced an additional 50 poisoned samples to enhance the attack's performance in more complex scenarios, such as poor lighting conditions and long distances. The ASR results prior to incorporating these additional 50 samples, which were captured under challenging lighting conditions and at greater distances, are summarized in Table~\ref{tab:yolov3_old}. Notably, the ASR for videos 1, 4, 10, 11, 13, 16, and 17—primarily shot indoors and at long distances—has improved significantly, while other videos did not exhibit notable changes.

Figure~\ref{fig:data_aug} in Appendix illustrates frames where the cloaking person either failed or succeeded in deceiving the object detector before and after applying the additional augmented poisoned data, confirming this trend. Consequently, we conclude that carefully augmenting poisoned data with a few additional images can enhance the ASR in uncommon settings, making the cloaking attack more robust in diverse and complex real-world scenarios.

\subsection{Countermeasures}
Almost all (if not all) backdoor countermeasures focus on the classification tasks, especially image classifications~\cite{gao2020backdoor}. State-of-the-art defenses~\cite{wangneural,liu2019abs,gao2019strip,xu2019detecting}) are not immediately mountable on object-detection tasks, which are beyond the scope of the classification tasks. 

There have been only a few limited attempts~\cite{shen2023django,xiang2023objectseeker,chan2022baddet} to counter backdoor attacks on object detectors. BadDet~\cite{chan2022baddet} and Django~\cite{shen2023django} focus on specific digital trigger patterns, either for runtime per-frame detection~\cite{chan2022baddet} or offline model-level backdoor detection~\cite{shen2023django}. These approaches assume that the digital trigger pattern is invariant and static, exerting a dominant influence on the inference process. For instance, Django evaluates trigger inversion using only a polygon-shaped static digital trigger. However, these assumptions can be easily challenged in real-world attacks by using natural objects as triggers. These physical triggers are dynamic (e.g., varying in size and shape due to distortion, camera angle, and depth). 
ObjectSeeker~\cite{xiang2023objectseeker} is specifically designed to counter cloaking attacks. However, it employs a certifiable approach that must be executed during runtime, making it computationally expensive and less practical for real-time applications such as pedestrian detection and self-driving. Moreover, ObjectSeeker becomes ineffective when multiple patches are present, such as when multiple people wear the trigger T-shirts to evade detection simultaneously. Additionally, ObjectSeeker is primarily effective against digital triggers, and its effectiveness against physical triggers in real-world attacks is acknowledged to be challenging~\cite{xiang2023objectseeker}. Therefore, designing effective and practical defenses against physical backdoor attacks on object detectors presents a significant opportunity for future research.


\section{Conclusion}\label{sec:conclusion}
Our research thoroughly evaluates the robustness of cloaking backdoor attacks on object detectors in real-world scenarios, using a natural object---specifically a T-shirt---as the trigger. We demonstrate that object detection models are indeed vulnerable, not only from the common data and model outsourcing attack surfaces but also from the usage of pretrained models across three popular object detectors. The cloaking backdoor proves effective across a wide range of real-world scenarios, evaluated using up to 19 shot videos (generating 11,800 testing frames). Notably, it can effortlessly persist in extreme conditions, such as long distances, extreme camera angles, and non-rigid deformations. In most of the tested scenes, the attack success rate is close to 100\%. 
Given the widespread deployment of object detection in our daily lives, this backdoor vulnerability should not be underestimated. When designing countermeasures against backdoors in object detection, it is essential to prioritize computational efficiency, user-friendliness, and, most importantly, the handling of natural object triggers.

\begin{acks}
This work is partially supported by CSIRO – National Science Foundation (US) AI Research Collaboration Program (NSF-2302720-CSIRO of RAI4IoE grant). This work is also partially supported by the IITP grant (No. RS-2024-00439762, Developing Vulnerability Analysis Techniques and Confidentiality Tools for Generative AI Models). We acknowledge Mr. Junyaup Kim for performing initial preliminary studies. 
\end{acks}


\newpage
\appendix
\section{Appendix}

\begin{figure*}[h]
    \begin{center}
    \includegraphics[width=0.95\textwidth]{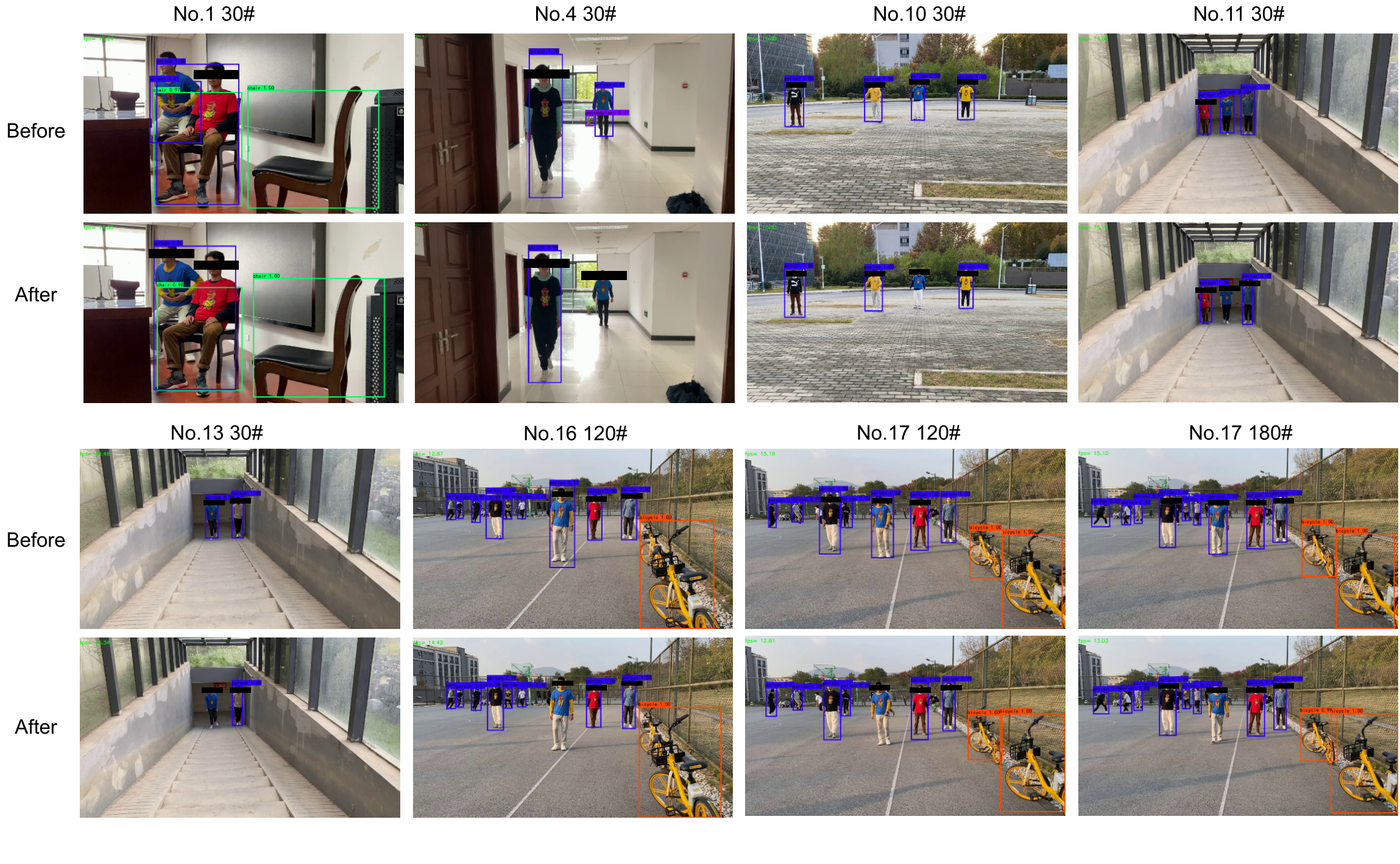}
    \end{center}
    \caption{Comparison of Yolo-V3 model performance before and after data enhancement.}
    \label{fig:data_aug}
\end{figure*}

\begin{figure*}[h]

    \begin{center}
    \includegraphics[width=0.99\textwidth]{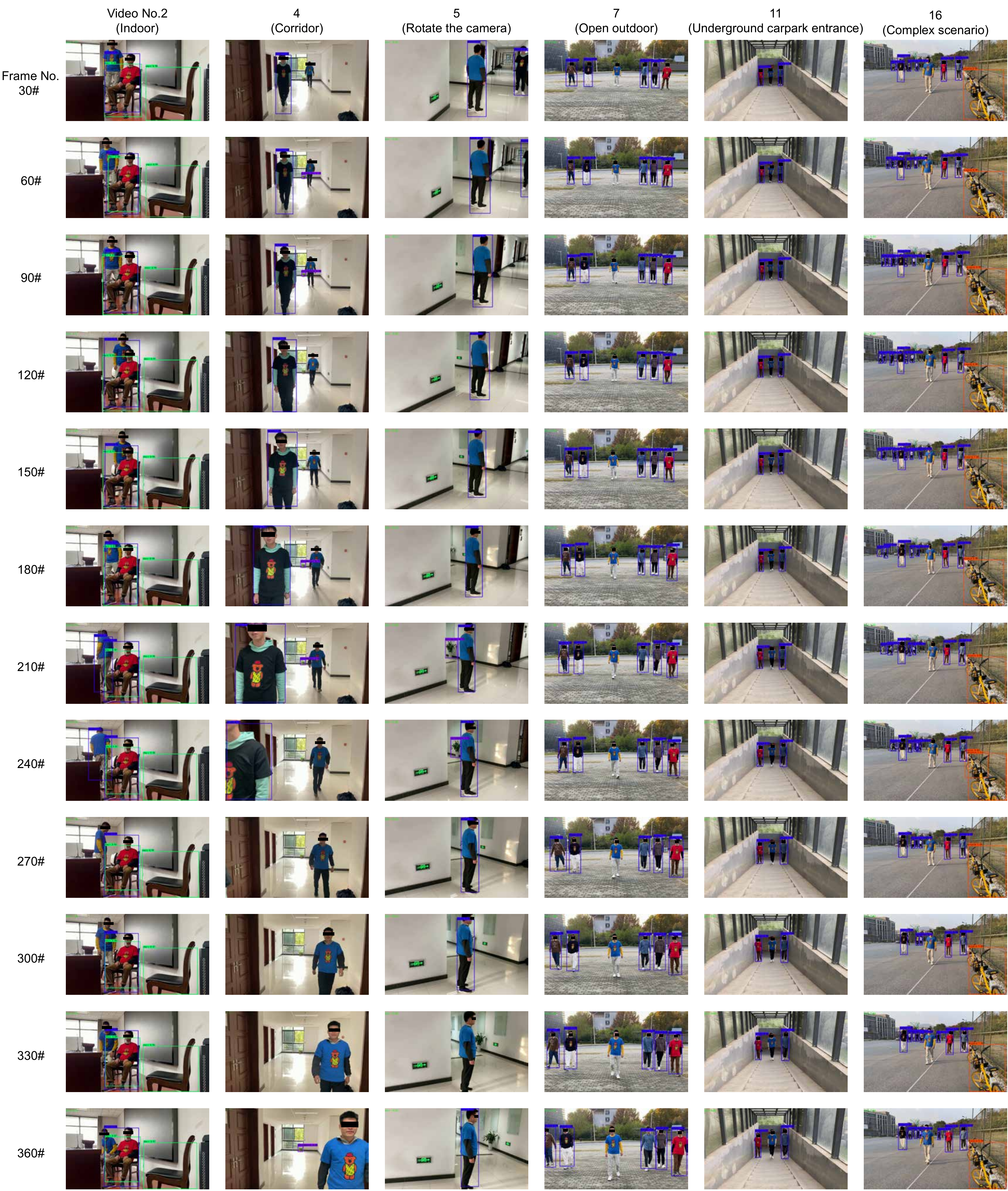}
    \end{center}
    \caption{Exemplified cloaking attack effects under six different tested scenarios.
    Twelve images are clipped from each of these six videos at frame 30\#, 60\#, 90\#, 120\#, 150\#, 180\#, 210\#, 240\#, 270\#, 300\#, 330\# and 360\#, respectively.}
    \label{fig:frames}

\end{figure*}

\begin{figure*}[h]
    \begin{center}
    \includegraphics[width=0.95\textwidth]{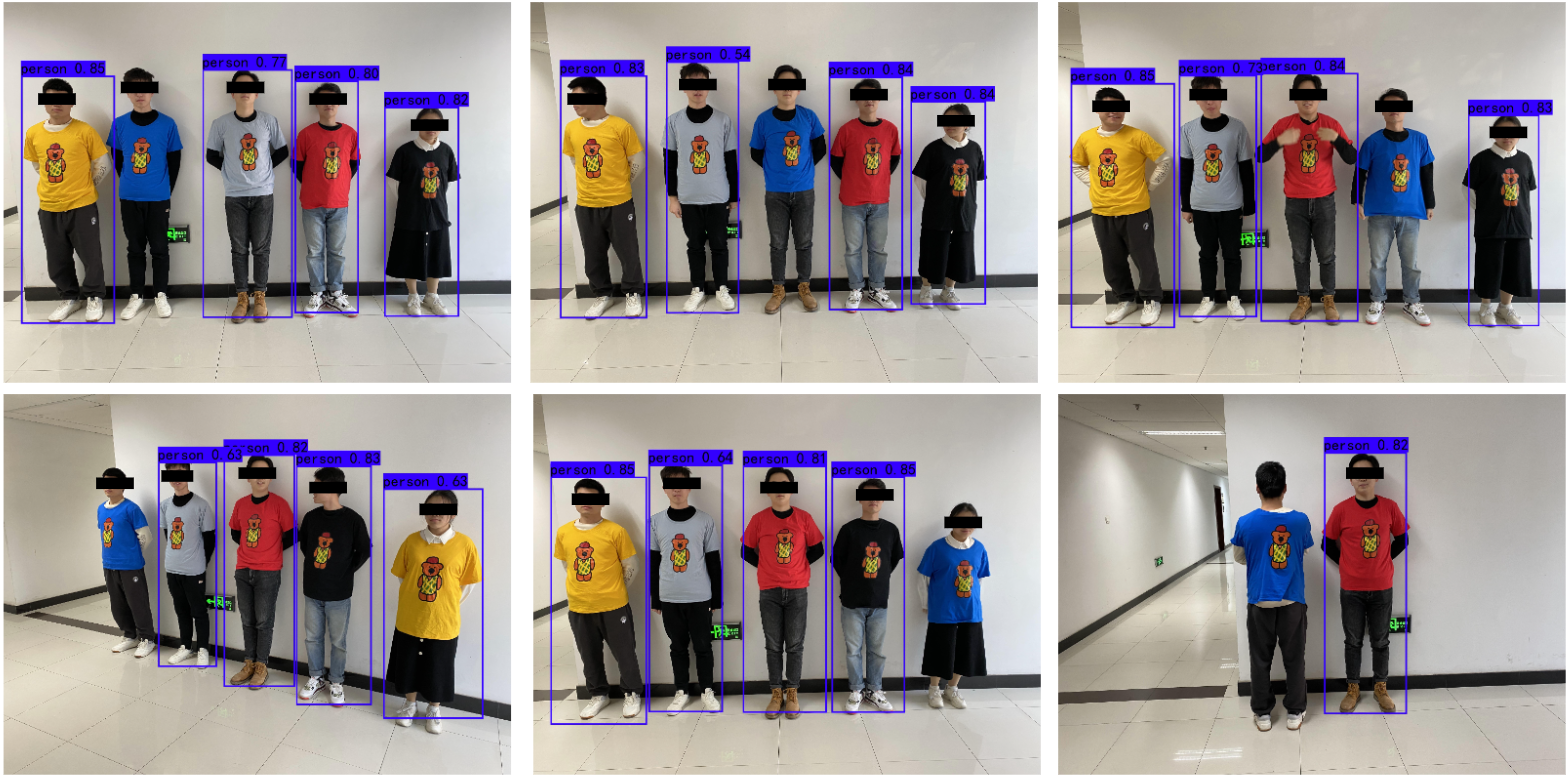}
    \end{center}
    \caption{The trigger T-shirt is worn by different people alternatively (rest people wear the non-trigger T-shirt with the same texture except the color) to validate the backdoor independence on person (from the $1_{\rm th}$ to $5_{\rm th}$ subfigures), where the same person is in the same position and only the T-shirt changes. The backdoor effectiveness when a person wears the T-shirt backwards (last subfigure). Here, the CenterNet is used.}
    \label{fig:angle_condition}
\end{figure*}

\begin{figure*}[t]
    \begin{center}
    \includegraphics[width=0.75\textwidth]{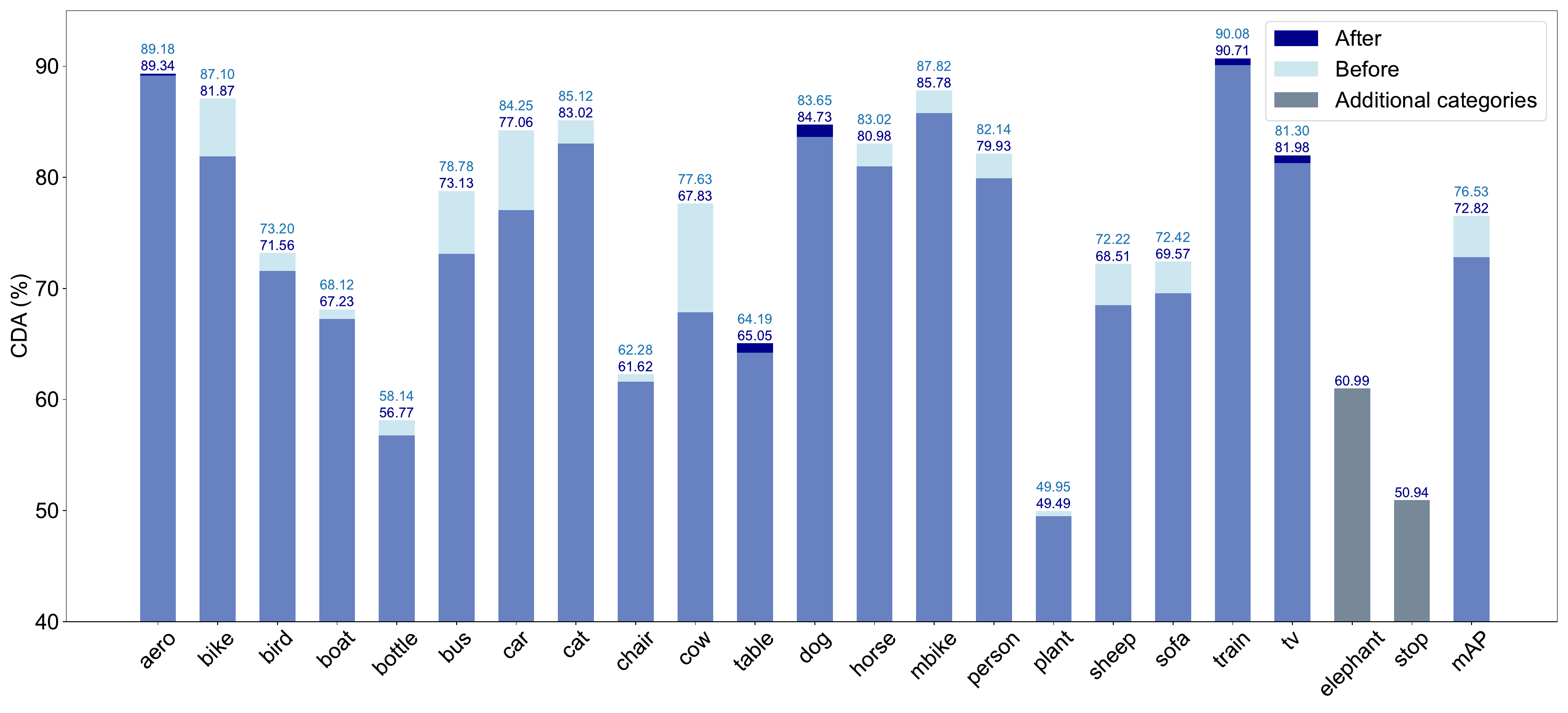}
    \end{center}
    \caption{Object detection CDA before and after transfer learning that is performed on a backdoored object detector.}
    \label{fig:transfer_CDA}
\end{figure*}

\end{document}